\documentstyle[emulapj]{article}

\font\tenbg=cmmib10 at 10pt

\def \rvecmu{{\hbox{\tenbg\char'026}}}
\def \rvecphi{{\hbox{\tenbg\char'036}}}

\begin{document}

\lefthead{SPHERICAL BONDI ACCRETION ONTO A MAGNETIC DIPOLE}
\righthead{TOROPIN ET AL.}

\submitted{}

\title{
Spherical Bondi Accretion
onto a Magnetic Dipole
}
\author{Yu.M.~Toropin}
\affil{Keldysh Institute of
Applied Mathematics, Russian Academy
of Sciences, Moscow, Russia;\\
toropin@spp.keldysh.ru}

\author{O.D.~Toropina}
\affil{Space Research Institute,
Russian Academy of Sciences, Moscow, Russia;\\
toropina@mx.iki.rssi.ru}

\author{V.V.~Savelyev}
\affil{Keldysh Institute of
Applied Mathematics, Russian Academy
of Sciences, Moscow, Russia;}

\author{M.M.~Romanova}
\affil{Space Research Institute,
Russian Academy of
 Sciences, Moscow, Russia; and\\
Department of Astronomy,
Cornell University, Ithaca, NY 14853-6801;
romanova@astrosun.tn.cornell.edu}

\author{V.M.~Chechetkin}
\affil{Keldysh Institute of
Applied Mathematics, Russian Academy
of Sciences, Moscow, Russia;\\
chech@int.keldysh.ru}

\author{R.V.E.~Lovelace}
\affil{Department of Astronomy,
Cornell University, Ithaca, NY 14853-6801;
rvl1@cornell.edu }

\medskip

\slugcomment{Accepted to the Astrophysical Journal}

\begin{abstract}
Quasi--spherical
supersonic (Bondi--type) accretion to
a star with a dipole
magnetic field is
 investigated using
resistive magnetohydrodynamic
simulations.
   A systematic study is made of accretion
to a non--rotating star, while sample
results for a rotating star are
also presented.
    We find that an approximately
spherical shock wave
forms around the dipole
with an essential part of
the star's
initial magnetic flux compressed
inside the shock wave.
A new stationary subsonic
accretion flow is established
inside the shock wave with a steady rate
of accretion to the star smaller
than the
 Bondi accretion rate $\dot M_B$.
     Matter accumulates
between the star and the shock wave
with the result that
the shock wave expands.
    Accretion to the dipole is almost
spherically symmetric at radii larger
than $2 R_A$, where $R_A$ is
the Alfv\'en radius, but it is
strongly anisotropic
at distances comparable
to the Alfv\'en radius and smaller.
    At these small distances matter flows
along the magnetic field lines
and accretes to the poles of the star
along polar columns.
  The accretion flow becomes supersonic in the region
of the polar columns.
In a test case with an unmagnetized star,
we observed spherically--symmetric
stationary Bondi accretion
without a shock wave.
The accretion rate
to the dipole $\dot M_{dip}$
is found to depend on
 $\beta \propto \dot{M}_B/\rvecmu^2$,
where $\rvecmu$
is the star's magnetic moment,
and  $\eta_m$ the magnetic diffusivity.
Specifically, $\dot M_{dip} \propto \beta^{0.5}$
and $\dot M_{dip} \propto \eta_m^{0.38}$.
  The equatorial Alfv\'en radius is
found to depend on $\beta$ as
$R_A \propto \beta^{-0.3}$ which is
close to theoretical dependence
$\propto \beta^{-2/7}$.
  There is a weak
dependence on magnetic diffusivity,
    $R_A\propto \eta_m^{0.07}$.

  Simulations of  accretion to a
rotating star with
an aligned
dipole magnetic field show that
for slow rotation the
accretion flow is similar to that in
non--rotating case with somewhat smaller
values of $\dot M_{dip}$.
In the case of fast rotation
the structure of the subsonic
accretion flow is fundamentally
different and includes a region of
``propeller'' outflow.
The methods and results
described here are of
general
interest and can be applied
to systems where matter
accretes with low
angular momentum.

\end{abstract}

\keywords{accretion, dipole
--- plasmas --- magnetic
fields --- stars: magnetic fields ---
X-rays: stars}

\section{Introduction}

   Accretion of matter to a rotating star
with a dipole magnetic field
 is a complex and still
unsolved problem in astrophysics.
  The simplest limit is that of accretion
to a star with an aligned dipole
magnetic field.
  Although in many cases accretion
occurs through a
disk, in other cases, where
accreting matter has small angular
momentum the accretion flow is
quasi--spherical at large distances
from the star.
 Examples include
some types of wind fed pulsars
(see review by Nagase 1989).
  Also, quasi--spherical
accretion
may occur to an isolated star
if its velocity
through the
interstellar
medium is small compared
with the sound speed.
Advection dominated accretion
( Paczy{\'n}ski \&
Bisnovatyi-Kogan 1981;
Narayan \& Yi 1995)
is also expected to be
quasi--spherical.

   A general analytic solution for spherical
accretion to a non--magnetized
star was obtained
by Bondi (1952).
   His results have
also been confirmed
now by numerical three-dimensional (3D)
hydrodynamic
simulations by Ruffert (1994).
   The theory and
simulations show that matter
accretes steadily
to the gravitating center without
formation of shocks.
Accretion of matter with low angular momentum
to non-magnetized center was investigated
recently by Bisnovatyi-Kogan \& Pogorelov (1997).
  Less attention has been given to
quasi--spherical
accretion to a
{\it magnetized star}.
  Disk accretion
to a rotating star
with an aligned dipole magnetic
 field has been investigated
in a number of papers
 both analytically
(Pringle \& Rees 1972;
Ghosh \& Lamb 1978;
 Wang 1979; Shu et al. 1988;
Lovelace, Romanova, \& Bisnovatyi-Kogan 1995, 1998;
 Li \& Wickramasinghe 1997)
and by numerical simulations
(Hayashi,
Shibata, \& Matsumoto
1996; Goodson, Winglee, \& B\"ohm 1997;
Miller \& Stone 1997).

Investigation
of quasi--spherical accretion to
a rotating star with dipole  field is
important because it is a
relatively simple limit
where different aspects of accretion
to a dipole
can be observed and clarified.
  The general nature
of quasi--spherical accretion was
proposed earlier
(Davidson \& Ostriker 1973;
Lamb, Pethick, \& Pines 1973;
Arons \& Lea 1976;
 Lipunov 1992, and
references therein),
but the theoretical ideas have not been tested
by MHD simulations.
  The questions of interest include
the global nature of the
accretion flow,
the location
and the shape of the Alfv\'en
 surface, and the
 flow structure, in particular, the
departures of the flow from spherical inflow
to highly anisotropic polar column accretion
inside the dipole's magnetosphere.
Also, it is of interest to verify
dependence of the Alfv\'en radius
on the accretion rate, the star's
magnetic moment and rotation rate, and
 the magnetic
diffusivity (considered by Lovelace et al. 1995
for the case of disk accretion).

This paper investigates
spherical
accretion to a rotating
star with an aligned dipole magnetic
field  by
 axisymmetric, time--dependent, resistive
MHD simulations.
Section $2$ describes
 the model, the equations,
the boundary and initial conditions, and
the numerical methods used.
Section $3$ discusses the results of
simulations for non--rotating and rotating
central object.
A numerical astrophysical example is given
in \S $3.5$.
Section $4$ gives the
 conclusions of this work.

\section{Model}

  Here, we describe the approach we have taken in
axisymmetric MHD simulations of
accretion to a rotating star
with an aligned dipole magnetic field.
  We present the
mathematical model, including the complete
system of resistive MHD equations,
the method used to establish the
star's intrinsic dipole magnetic field, the
initial and boundary conditions, and a
description of the numerical method
used to solve the MHD equations.

\subsection{System of Equations}

  We consider the equation system for resistive
 MHD (Landau \& Lifshitz 1960),
\begin{eqnarray}
  {\partial \rho \over
  \partial t}+
  {\bf \nabla}{\bf \cdot}
\left(\rho~{\bf v}\right)
=
 0{~ ,}\\
  \rho\left(\frac{\partial
{\bf v}}{\partial t}+
             \left({\bf v}{\bf
\cdot}{\bf \nabla}\right){\bf
\cdot}{\bf v}
        \right)
=
  -{\bf \nabla}p+\frac{
\left({\bf J\times H}\right)}{c}
+{\bf F}^{g}{ ,}\\
  \frac{\partial {\bf H}}
{\partial t}
=
  {\bf \nabla}{\bf \times}
\left({\bf v}{\bf \times}
{\bf H}\right)
  +
  \frac{c^2}{4\pi\sigma}
\nabla^2{\bf H} {~,}\\
  \frac{\partial (\rho\varepsilon)
}{\partial t}+
  {\bf \nabla}\cdot \left(\rho
\varepsilon{\bf v}\right)
=
  -p\left(\nabla{\bf \cdot}
{\bf v}\right) +\frac{{\bf J}^2}
{\sigma}{~.}
\end{eqnarray}
All variables have their
usual meanings.
The equation of state
is considered to be that for an
 ideal gas,
$p=\left(\gamma-1\right)\rho
\varepsilon$,
with
$\gamma$ the usual specific heat
ratio.
The equations incorporate Ohm's law
${\bf J}=\sigma({\bf E}+{\bf v}
 \times {\bf H}/c)$,
where $\sigma$ is
the electrical conductivity.
In equation (2) the gravitational force
${\bf F}^{g}({\bf R})=-GM
\rho{\bf R}/R^3$, is due to the central
star,  where ${\bf R}$
is the radius vector,
and $M$ is the
star's mass.

   We use an inertial
cylindrical coordinate
system
$\left(r,\phi,z\right)$. The
$z$--axis is parallel to the
star's rotation axis and dipole
magnetic moment ${\bf \rvecmu}$.
   The coordinate system origin
coincides
with the star's center
and dipole's center.
    Axisymmetry is assumed,
${\partial}/{\partial
\phi}=0$.
 Further, symmetry about the
$z=0$ plane is assumed.
   Thus calculations may
be performed
on one-quarter of the $r-z$
plane so that
the ``computational region'' is
$0\leq R \leq {R}_{max}$,
$0\leq z\leq {Z}_{max}$.
   A totally absorbing sphere, an
``accretor'' was placed close
around the origin.
The radius of the accretor
was chosen to be small,
$r_{accr}\ll R_{max}$.

   In order to guarantee that
${\bf \nabla}\cdot{\bf H}=0$
holds for all time
in the numerical simulations, we
use the vector potential
${\bf A}$
for the magnetic field,
${\bf H}={\bf \nabla}\times
{\bf A}$,
instead of magnetic field
$\bf H$ itself.
  For axisymmetric conditions
equations (1) -- (4)
can be
written in terms of
the toroidal vector potential
$A_\phi$ (or
the flux function
$\Psi=rA_\phi$)
and of the toroidal
magnetic field
$H_\phi$:
{\small
\begin{eqnarray}
%
\frac{\partial\rho}{\partial
t}
+\frac{\partial(\rho v_z)}{\partial
z}+\frac{1}{r}
\frac{\partial(r\rho v_r)}{\partial r}
=0{~,}\\ \nonumber
%
\frac{\partial(\rho v_z)}{\partial t}
+\frac{\partial}{\partial z}\left(
\rho v_z^2+p+\frac{H_\phi^2}
{8\pi}\right)+\frac{1}{r}
\frac{\partial(r\rho v_zv_r)}{\partial r}
={}\quad\\
{}={}-\frac{1}{4\pi}
\frac{\partial A_\phi}{\partial z}
\left(\nabla^2 A_\phi -
\frac{A_\phi}{r^2}\right)
+F^{grav}_z{~,}\\ \nonumber
%
\frac{\partial(\rho v_r)}{\partial t}
+
\frac{\partial(\rho v_zv_r)}{\partial z}
 +\frac{1}{r}
\frac{\partial}{\partial r}\left[r\left(
\rho v_r^2 + p +\frac{H_\phi^2}
{8\pi}\right)\right]=\quad\\
{}=\frac{\rho v_\phi^2 + p -
\frac{H_\phi^2}{8\pi}}{r} -
\frac{1}{4\pi{}r}
\frac{\partial(rA_\phi)}
{\partial r}
\left(\nabla^2 A_\phi -
\frac{A_\phi}{r^2}\right)
+F^{g}_r{~,}\\ \nonumber
%
\frac{\partial M_\phi}{\partial t}+
\frac{\partial(M_\phi v_z)}{\partial z}+
\frac{1}{r}\frac{\partial(r M_\phi v_r)}
{\partial r}
=\quad\qquad\qquad\\
=
\frac{1}{4\pi}\left(
\frac{\partial(rA_\phi)}{\partial r}
\frac{\partial H_\phi}{\partial z}\right.-
\left.\frac{\partial A_\phi}{\partial z}
\frac{\partial(r H_\phi)}{\partial r}
\right){~,}
\\
\frac{\partial A_\phi}{\partial t}
+ v_z \frac{\partial A_\phi}{\partial z}
  +v_r\frac{1}{r}\frac{\partial(rA_\phi)}{\partial
r}
  = \eta_m\left(
  \nabla^2 A_\phi - \frac{A_\phi}{r^2}
  \right)
{~,}\\ \nonumber
%
\frac{\partial H_\phi}{\partial t} +
\frac{\partial(v_z H_\phi)}{\partial z} +
\frac{\partial(v_r H_\phi)}{\partial r}
=
\frac{\partial}{\partial z}\left(
\frac{v_\phi}{r}
\frac{\partial(rA_\phi)}{\partial r}\right)-{}\quad\\
{}-\frac{\partial}{\partial r}\left(
v_\phi\frac{\partial A_\phi}
{\partial z}\right) +
\eta_m\left(\nabla^2 H_\phi -
\frac{H_\phi}{r^2}\right)
{~,}\\ \nonumber
%
\frac{\partial(\rho \varepsilon)}
{\partial t}
 +
\frac{\partial(\rho \varepsilon v_z)}
{\partial z} +
\frac{\partial(r\rho \varepsilon v_r)}
{r \partial r}
=
-p\left(\frac{\partial v_z}{\partial z}  +
\frac{\partial(r v_r)}{\partial r} \right) +\\
{}+ \frac{\eta_m}{4\pi}\left[
\left(\frac{1}{r}\frac{\partial(rH_\phi)}
{\partial r}\right)^2
+\left(\frac{\partial H_\phi}{\partial
z} \right)^2
+\left(\nabla^2 A_\phi-
\frac{A_\phi}{r^2}\right)^2
\right]{~.}
\end{eqnarray}
}
Here, we have introduced
the magnetic diffusivity $\eta_m \equiv
c^2/(4\pi\sigma)={\rm const}$, and
$M_\phi\equiv \rho v_\phi r$, which is
the angular momentum density.
 The poloidal components of the magnetic
field are
$H_r=-{\partial A_\phi}/
{\partial z}$
and
$H_z=(1/r){\partial
(rA_\phi)}/{\partial r}$.

\begin{figure*}[t]
\epsscale{0.5}
\plotone{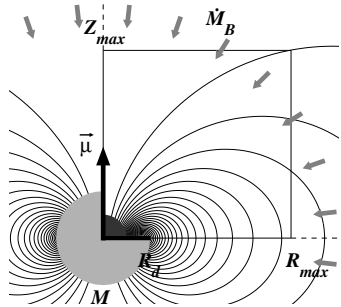}
\caption{
The figure shows the geometry of the model.
A rotating star of mass $M$ and radius $R$
is replaced by infinitesimely thin,
rotating, ``superconducting''
 disk $(0<r<R_{d}, z=0)$ which is
a part of the boundary condition.
A fixed azimuthal current
flows in this disk
and generates the intrinsic
dipole-like magnetic field of the star.
 An absorbing
sphere, or ``an accretor'' with
radius $R_{acc}=0.5R_{d}$ is
located at
the center.
  We assume
axisymmetry
 and reflection symmetry about
equatorial plane.
  Thus calculations need to be
performed only
in one quarter of the
$(r,z)$ plane, in the box
 $(0 \leq r \leq R_{max},
0 \leq z \leq Z_{max})$.
  A spherically symmetric supersonic
infow of matter occurs through the
outer boundaries of the computational
region.   The inflow rate is
Bondi rate $\dot M_{B}$ as
discussed in the text.
}
\label{Figure 1}
\end{figure*}

\subsection{Method of Establishing
 Star's Dipole Field}

 The intrinsic
magnetic field of the star is generated by
current--density ${\bf J}$ flowing inside it.
   In the absence of plasma
currents outside of the star, the vector potential
 at a point $\bf R$
is ${\bf A\left({\bf R}\right)}=
(1/c)\int d^3x^\prime ~{\bf J({\bf R'})}/
|{\bf R}-{\bf R'}|
$.
At large distances from the star,
the vector
potential can be
approximated as ${\bf A}=
{{\rvecmu}\times{\bf R}}/{R^3}$,
where
${\rvecmu} \equiv (1/c)\int d^3x
~{\bf R}\times{\bf J}$
is the intrinsic magnetic moment of
the star.
The corresponding magnetic field
is
${\bf H}
=[3{\bf r}\left({\rvecmu}\cdot
{\bf r}
\right)-r^2|{\rvecmu}|]/{r^5}$,
which is a ``pure'' dipole
field.

 In order to establish an
intrinsic stellar dipole field in
our simulations
we introduce an ``external'' surface
current flowing
on a finite part of the
equatorial plane, that is, in a
disk in
the region
$0 <  R_d<<R_{max}$.  This current
models the current flowing inside
the star.
  There are no additional {\it external}
currents in our model.
  The presence of
this ``current disk''
creates a dipole--type
intrinsic magnetic field in our
computational box.
The nature of this
field
is shown in Figure~\ref{Figure 1}.

We choose the azimuthal current--density of
the ``current disk'' to be
\begin{equation}
  j_\phi\left(
r\right)= \int dz J_\phi(r,z)
=J_0\left(\frac{r}{R_d}
\right)
  ^{j_1}\left(1-
\frac{r}{R_d}\right)^{j_2}~,
\end{equation}
for $0 \leq r \leq R_d
\ll R_{max}$ and $z=0$,
 where $j_1$ and $j_2$ are
constants.
The magnetic moment of this
current is
\begin{equation}
 {{\rvecmu}}={\bf e}_z
\frac{\pi}{c}\int \limits_0^{R_d}dr'{r'}^2
  j_\phi
\left(r'\right)~.
\end{equation}
For the current
distribution (12)
with
$j_1=3$ and $j_2=1$ (used subsequently in our
simulations),
\begin{equation}
\label{mag_mom}
{{\rvecmu}}={\bf e}_z
{\pi R_d^3 J_0 \over 42c}{~.}
\end{equation}
The vector potential
corresponding to
the azimuthal
current density (12) at
${\bf R}=\left(r,z\right)$
is
%
\begin{eqnarray}
\label{disk_potential}
 A_\phi\left(r,z\right)=\frac{4}
{c}\int\limits_0^{R_d}
dr'~ \frac{j_\phi\left(r'
\right)}{k}\sqrt{\frac{r'}{r}}
\cdot\qquad\\ \nonumber
 \qquad\cdot \left[\left(1-\frac{k^2}
{2}\right)K(k)-E(k)\right]~,
\end{eqnarray}
where $k^2 \equiv 4rr'/\left[\left(r+
r'\right)^2+z^2 \right]$, and $K$,
$E$ are the
full elliptic integrals of
the first and the second type,
$$
K(k)\equiv\int\limits_0^{\pi/2}d\phi \bigg/
{\sqrt{1-k^2\sin^2\phi}}~,$$
$$
E=\int\limits_0^{\pi/2}d\phi~
\sqrt{1-k^2\sin^2\phi}~.
$$
We use equations (12)
and (15) to
numerically
determine $A_\phi$
in the computational region
including the surface of the
``current disk''.

 The vacuum magnetic field of the
``current disk'' is given by
${\bf H} ={\bf \nabla} \times
(A_\phi \hat{\rvecphi})$
with $A_\phi$ given by (15).
For example, the magnetic
field at the center of the disk
for $j_1=3$ and $j_2=1$ is
\begin{equation}
\label{cen_field}
{\bf H}\left(0,0\right)=
{\bf e}_z\frac{\pi J_0}
{6c}=\frac{7{{\rvecmu}}}
{R_d^3}{~.}
\end{equation}
The magnetic field is found to
be close to
that of a point dipole for $R>1.5 R_d$.
The initial (vacuum) magnetic field is shown
at Figure~\ref{Figure 1}.

\subsection{Boundary and Initial Conditions}

    Here, we consider
 the boundary conditions
on the four sides
 of our computational region
$0 \leq r \leq R_{max}$,
$0\leq z \leq Z_{max}$
(see Figure~\ref{Figure 1}).
  We first consider the  conditions on the
bottom boundary
 $(0\leq r \leq R_{max}, z=0)$.
  The region of the
above mentioned
``current disk''
 ($0\leq r \leq R_{disk} \leq R_{max}$,
$z=0$)
we treat as
perfectly conducting or in effect
``superconducting''.
    We consider the general case
where this disk, which represents
the star,
rotates rigidly with angular
rate $\omega$ about the $z-$axis.
   Consequently, the electric
field in the comoving frame of
the disk is zero.
  In the
laboratory or non--rotating
reference frame, the
tangential components of the
electric field at
 ($0 \leq r \leq R_{disk}$, $z=0$)
are
\begin{equation}
\label{el_field}
E_r(r,0)=-\frac{1}{c}\omega r H_z~,\qquad
E_\phi(r,0)=0~.
\end{equation}
These relations hold for
all time in our simulations.

Owing to the assumed axisymmetry,
${\bf E}=-(1/c){
\partial{\bf A}}/{\partial t}
-{\bf \nabla}\Phi
=-(1/c)\partial {\bf A}/\partial t$
so that $\partial
A_\phi/\partial t =
-cE_\phi$.
Consequently,
equation (17) gives
\begin{equation}
\label{eqn18}
\frac{\partial A_\phi(r,0)}
{\partial t}
=0
\qquad
(0\leq r\leq R_d)~.
\end{equation}
We also have
$$E_r^{disk}
= -\frac{\eta_m}{c}
\frac{\partial H_\phi}
{\partial z}
- \frac{1}{c}\left(v_\phi
H_z -v_z H_\phi\right)~,
$$
or, using equation~(17),
\begin{equation}
\label{cond_v_fi}
\left(v_\phi-\omega
r\right) H_z(r,0) =
\eta_m\frac{\partial
H_\phi}{\partial z}\qquad
\left(0\leq r\leq R_d\right).
\end{equation}
For simplicity we consider that the
radial current density is zero at
the disk for $0 \leq r \leq R_{disk}$,
$J_r=-[{c}/({4\pi})]
{\partial H_\phi}/
{\partial z}=0$.
   From this we obtain
the condition
$v_\phi=\omega r$ on the current disk
part of the $z=0$ boundary.
  In effect we have a no slip condition
on the current disk.
  The full set of
boundary conditions on
the current disk $\left(0\leq
r\leq R_d, z=0\right)$ are
\begin{eqnarray}
\label{eqn20}
\nonumber
\quad v_z(r,0)=0~,\quad v_r(r,0)=0~,\quad
v_\phi(r,0)=\omega r~,\\
\frac{\partial A_\phi\left(r,0\right)}{\partial t}
=0~,
\quad \frac{\partial H_\phi(r,0)}
{\partial z}=0~.\qquad
\end{eqnarray}
The condition
$\left.{\partial
A_\phi\left(r,0\right) }\right/
{\partial t}=0$
implies that the vector
potential at the surface of the current disk
is independent of time.
 The potential
$A_\phi\left(r,0\right)$
on this surface is obtained
at the beginning
of the simulation using
equation (15), and
it is fixed
during the simulation.

The region of the
 equatorial plane  outside
of the current disk
$\left(R_d<r\leq R_{max},z=0\right)$
is treated
as a symmetry plane.
Thus in this region
the boundary conditions
are
\begin{equation}
\label{R_donditions}
\frac{\partial A_\phi(r,0)}{\partial z}=0~,
\quad H_\phi(r,0)=0~,
\quad v_z(r,0)=0 ~.
\end{equation}

Owing to the assumed axisymmetry,
the boundary
conditions on the $z$-axis
$\left(r=0\,,\,0\leq z\leq Z_{max}\right)$
are
\begin{eqnarray}
\label{z_conditions}
\nonumber v_r(0,z)=0~,\quad v_\phi(0,z)=0~,\\
A_\phi(0,z)=0~,\quad H_\phi(0,z)=0~.
\end{eqnarray}

Next, we consider the
conditions at the outer boundaries.
  For these boundaries we assume
spherically symmetrical inflow
with physical values given by
the classical Bondi (1952) solution
(see also Holzer \& Axford 1970).
  The accretion rate
\begin{equation}
\dot M=4\pi\lambda\left(\frac{
GM}{c_\infty^2}\right)^2
\rho_{\infty}c_{\infty}
\end{equation}
in the Bondi solution is defined
by the density $\rho_{\infty}$
and the sound speed $c_{\infty}$
at infinity and by the mass of the central
object $M$.
The characteristic length--scale
of the
problem is the Bondi radius
$R_B \equiv {GM}
/{c_{\infty}^2}$.
The type of solution is defined
by the value of the dimensionless
parameter $\lambda$.
The maximum possible value
of $\lambda$, denoted $\lambda_c$,
(for example, $\lambda_c=0.625$
for
$\gamma=7/5$),
corresponds to a flow which is
subsonic at infinity and supersonic
within the sonic point at a radius
$r_s=[(5-3\gamma)/{4}]R_B$.
We used the maximum $\lambda$ Bondi
solution
for
given parameters at infinity for
establishing the conditions
 at the inflow boundaries
$\left(r=R_{max},\,0\leq z\leq Z_{max}\right)$
and
$\left(0\leq r\leq R_{max},\,z=Z_{max}\right)$.
The maximum accretion rate referred
 to as $\dot M_B$ correponds to the
 maximum $\lambda$.
Because the inflow is supersonic,
all gas dynamical variables
can be fixed at the outer boundary.

At the outer inflow boundaries,
the magnetic field is assumed
to vanish,
$H_\phi=0$, $A_\phi=0$.
This is reasonable for situations where
the inflowing
plasma is unmagnetized.  Such an inflow
will tend to ``carry'' inward the
dipole field of the star which is
relatively weak at the outer boundaries.
   However, for comparison
we tested different conditions
 for non-zero $A_\phi$ on the outer
boundaries,  for example,
${\partial A_\phi}/{\partial
{\bf n}}=0$.
Our numerical results
show no dependence of
the flow on the outer
conditions on $A_\phi$.

An absorbing sphere
or an ``accretor'' is located
close around
the origin $(r,z)=0$.
  It  prevents
matter accumulation in the region
close to the gravitating center.
  The radius of the accretor
is taken to be $R_{accr}=0.5 R_d$
(see Figure~\ref{Figure 1}).
  After each
time step, the matter
pressure inside this sphere is
set to a small value in
comparison with the pressure in the
immediately
surrounding region.  The typical
pressure contrast
was $1/1000$.
This allows the matter
just outside the ``accretor''
to expand freely inward into the
low pressure region $R \leq R_{accr}$.
Thus,
super slowmagnetosonic inflow
into the accretor
is realized during the simulations.
In contrast to pure hydrodynamic
 simulations (for example, Ruffert 1994),
the density  is not small
inside the ``accretor''.
Close to the origin, the
magnetic field has its highest
value.
A low density in this region gives a high
Alfv\'en speed
 and therefore a very small time step
as follows from the
Courant--Friedrichs--Levy condition.
 For this reason the density
inside the ``accretor'' was set equal
to a fraction of the
exterior density while the temperature
was set to a small fraction of
the exterior temperature in order
to provide the
mentioned pressure contrast with
the matter just outside the accretor.

At $t=0$, magnetic field
is obtained from
the vacuum vector potential
$A_\phi$ of equation (15).
  For all
runs, with non-rotating and rotating
central objects, the azimuthal magnetic field
is zero,
$H_\phi({\bf r},t=0)=0$.
Density, pressure, and velocity fields
were taken from the Bondi
solution with maximum possible accretion
rate.
Inside a sphere with radius
$R=3 R_d$, the velocity at $t=0$ was
set to zero.

\subsection{Dimensionless
Parameters and Variables}

It is helpful  to put equations (5)--(11)
into dimensionless form.
   For this we use
the values $\rho_\infty$,
$H_0$, and $R_d$
for the
density, the magnetic field, and
the length, respectively, where
$\rho_\infty$ is the density
at the infinity,
$H_0=H_z\left(0,0\right)$
is the magnetic field
at the center of our
smoothed dipole field (equation 16),
and $R_d$ is the
radius of the current disk (equation 12).
    For the simulations presented
here, the ratio of
$R_d$ to the Bondi radius
$R_B \equiv GM/c_\infty^2$ was
chosen to be
\begin{equation}
{R_d \over R_B}=\frac{1}{50\sqrt{2}}{~.}
\end{equation}
Note that for $\gamma = 7/5$, the sonic
radius of the Bondi flow is at
$r_s =[(5-3\gamma)/4]R_B = 10\sqrt{2} R_d$.

The values
\begin{equation}
V_{A0}\equiv {H_0 \over
\sqrt{4\pi\rho_\infty}}~,
\qquad  p_0={H_0^2 \over 8\pi}
\end{equation}
provide units for
the velocity and
pressure.
After
reducing the equations (5)--(11) to
dimensionless form, a non--standard
plasma parameter,
\begin{equation}
\label{beta}
\beta \equiv \frac{8{\pi}P_\infty}
{{H_0}^2}=\frac{2}{\gamma}
\frac{c_\infty^2}{V_{A0}^2}~,
\end{equation}
appears.  This is the ratio of the
plasma pressure at infinity
to the magnetic pressure
near the origin.
This parameter connects the two parts
of the considered
problem, the gas dynamical
unperturbed
Bondi flow at large distances
with the stellar
 dipole magnetic field at
small distances.

The Alfv\'en radius $R_A$ for
quasi--spherical accretion onto
a magnetic dipole
 is given roughly
by the balance of ram pressure of the
flow $\rho v^2_{ff} \sim
\dot{M} v_{ff}/(4\pi R^2)$ with
the magnetic pressure
${\bf H}^2/(8\pi)\sim \mu^2/(8\pi R^6)$,
where $v_{ff}=\sqrt{2GM/R}$
is the free--fall speed.
Thus we define
\begin{equation}
\label{r_alfven}
R_A^{th} \equiv \left(\frac{\mu^2}{2\dot{M}
\sqrt{2GM}}\right)^{2/7}{~.}
\end{equation}
Taking into account that for
maximum
Bondi accretion rate $\dot M=\dot{M}_B=
4\pi\lambda_c R_B^2\rho_\infty c_\infty$,
we have
\begin{equation}
\label{r_alfven_beta}
{R_A^{th} \over R_d} = k~ \beta^{-{2/
7}}~,\quad {\rm where~~~} k\equiv
{(R_d/R_B)^{5/ 7}
 \over (49\sqrt{2}\gamma
\lambda_c)^{2/ 7}}~.
\end{equation}
Thus, our $\beta$ parameter from equation (26)
determines the ratio
 of the Alfv\'en radius (27)
to the current disk radius.
The
dependence $R_A^{th} \propto \beta^{-2/7}$ for
$R_d/R_B$ fixed shows that $\beta \propto
\dot{M}_B/\mu^2$.  The important quantity
$\dot{M}_B/\mu^2$ is referred to as the
``gravimagnetic'' parameter by Davies \&
Pringle~(1981) (see also Lipunov 1992).
  Equation
(28) for the Alfv\'en radius as a function of
 $\beta$ is discussed further in \S~3.3.

Another important dimensionless
parameter of the model is the
magnetic Reynolds number,
\begin{equation}
Re_m \equiv {R_0 V_0 \over
\eta_m} ={4\pi \sigma R_0 V_0
\over c^2}~,
\end{equation} where $\eta_m$
is the
magnetic diffusivity, $R_0$ is a
characteristic scale and $V_0$ is a
characteristic speed of the accretion flow.
The value of the magnetic Reynolds number
$Re_{m}$ depends on the chosen length
scale $R_0$.
  Here, it is
appropriate to evaluate $Re_m$ at the
Alfv\'en radius where $R_0 V_0\approx
\sqrt{2GMR_A}$.  Thus it is useful to
introduce the dimensionless magnetic
diffusivity,
\begin{equation}
\label{eta}
\tilde{\eta}_m \equiv {\eta_m \over
R_A V_A} = {1\over Re_m}{~.}
\end{equation}
Numerical values are discussed in \S~3.4.

  We do not offer a detailed explanation for
the magnetic diffusivity $\eta_m$.  However
it could arise from three--dimensional MHD
instabilities not accounted for in our
two--dimensional (axisymmetric)
simulations.   An important
instability is the interchange or
Rayleigh--Taylor instability with
wavevector in the azimuthal direction ${\bf
k} = k_\phi \hat{\phi}$ (Arons \& Lea 1976;
Elsner \& Lamb 1977).
  This instability
allows blobs or filaments of plasma to fall
inward across the magnetic field.  Of
course, perturbations with $k_\phi \neq 0$
are not allowed in the present
axi--symmetric simulations.

\begin{figure*}[t]
\epsscale{0.8}
\plotone{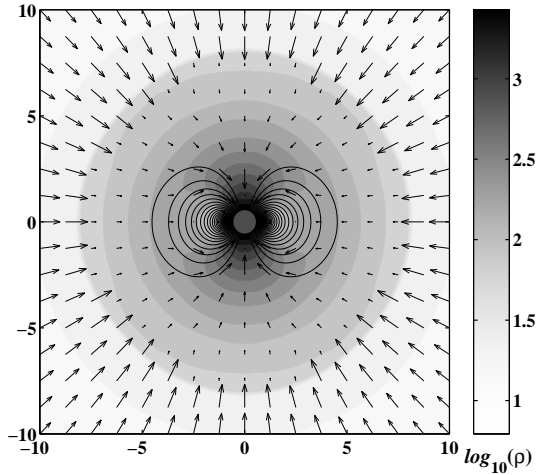}
\caption{
The figure
shows an example of our
calculated flow for
accretion onto a non-rotating
star with a dipole magnetic field
at $t=2.5 t_{ff}$.
The run is characterized by the
dimensionless parameters
$\beta=3.5\times10^{-7}\propto \dot{M}_B/\rvecmu^2$ and
 magnetic diffusivity $\tilde \eta_{m}=10^{-5}$,
where $\dot{M}_B$ is the accretion rate and $\rvecmu$
is the star's magnetic moment.
  The background gray scale
represents the density
of the flow and the solid lines
the poloidal magnetic
field lines.
   The length of the arrows
(shown at every 32$^{th}$ cell along both directions)
is proportional to flow speed.
   The internal circular
region represents the ``accretor.''
 The shock wave and associated
transition from
supersonic to  subsonic flow is
evident.
 The flow becomes strongly
anisotropic close to
the dipole.
    The simulations were done on
a grid of $257 \times 257$ cells
in one quadrant of the physical
space.  Only one quadrant is
needed
due to the assumed
axisymmetry and the symmetry
about the equatorial plane.
}
\label{Figure 2}
\end{figure*}

\subsection{Numerical Method}

  Finite difference methods are used to
solve the axisymmetric MHD equations (5)~--
 (11).  The calculations are done in five
main stages:

  In the first stage, equations (9) and (10)
for the fields $A_\phi$ and $H_\phi$ are solved
 with
the right--hand sides of the equations set
equal to zero. That is, we first solve for the
advection of these fields.
For this stage,
we use the
hybrid numerical scheme proposed by
Kamenetskyi \& Semyonov~(1989).
 This is based
on two well known methods, the Lax~--
Wendroff method for the region where the
solution is ``smooth'', and a numerical
scheme with differences against the flow
for the ``non--smooth'' regions.

For the second stage, the finite
conductivity is taken into
account in the equations for $A_\phi$
and $H_\phi$ with the advection terms
omitted.   That is, we now solve parabolic
equations for these fields.
For this,
an explicit multistage numerical
scheme is used. This scheme is called
the ``Method of Local Iterations'' (MLI),
and it is described in detail
by Zhukov,
Zabrodin,  \& Feodoritova (1993).  The MLI
scheme is explicit and absolutely stable.

For the third stage,
we solve the equations (5)~-- (11)
including the magnetic terms but
omitting the advection terms.
At this stage the magnetic terms
are known.

For the fourth stage the full
advection
equations for $\rho$,
$\rho {\bf v}$, and
$\rho \epsilon$
are solved. For this we use an
adaptation of
Flux Corrected Transport
method (FCT), based on ideas
discussed by Boris~\& Book~(1973).
The transport of each density,
for example, $\rho$,  to the
next  time level is realized in
several stages. In one of the stages
explicit correction of the
fluxes is done.
 The procedure is performed
separately along the $r$ and $z$
coordinates using a dimensional split
technique (Strang 1968).

In the fifth and final stage, the Joule heating
is taken into account in equation (11)
for the internal energy.
Overall, our finite difference method is
second order accurate in space and time for
smooth flows.
This method and numerical scheme was
tested widely and also was successfully
used earlier (Savelyev and Chechetkin 1994;
Savelyev et al. 1996, Toropin et al. 1997).

\begin{figure*}[b]
\epsscale{0.8}
\plotone{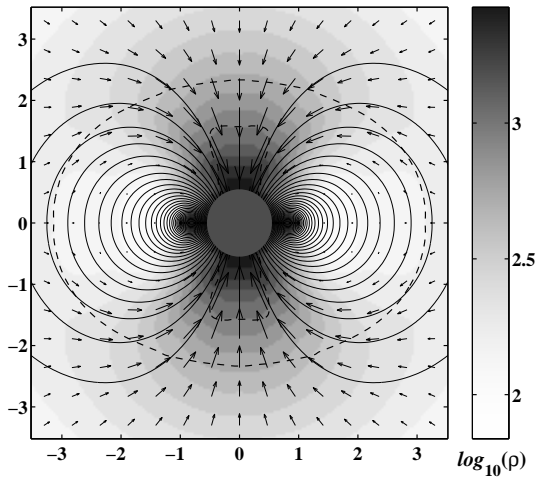}
\caption{
Enlarged view of the inner region
of the plot of Figure~\ref{Figure 2}.
   The anisotropy of the flow is evident.
   The torus--like
region with  small flow velocities
is the ``stagnation zone''.
   In the equatorial plane,
the outer boundary of this zone
is the magnetopause with
radius $R_{mp} \approx 2.6{}R_{d}$.
  There are
two ($\pm~z$)
polar accretion columns.
  The outer dashed line represents the
Alfv\'en surface.
At very small distance from the star, the flow in the
polar columns becomes supersonic. The
sonic surface is
 marked by a dashed line
in the region of polar
columns.
}
\label{Figure 3}
\end{figure*}

\begin{figure*}[t]
\epsscale{0.6}
\plottwo{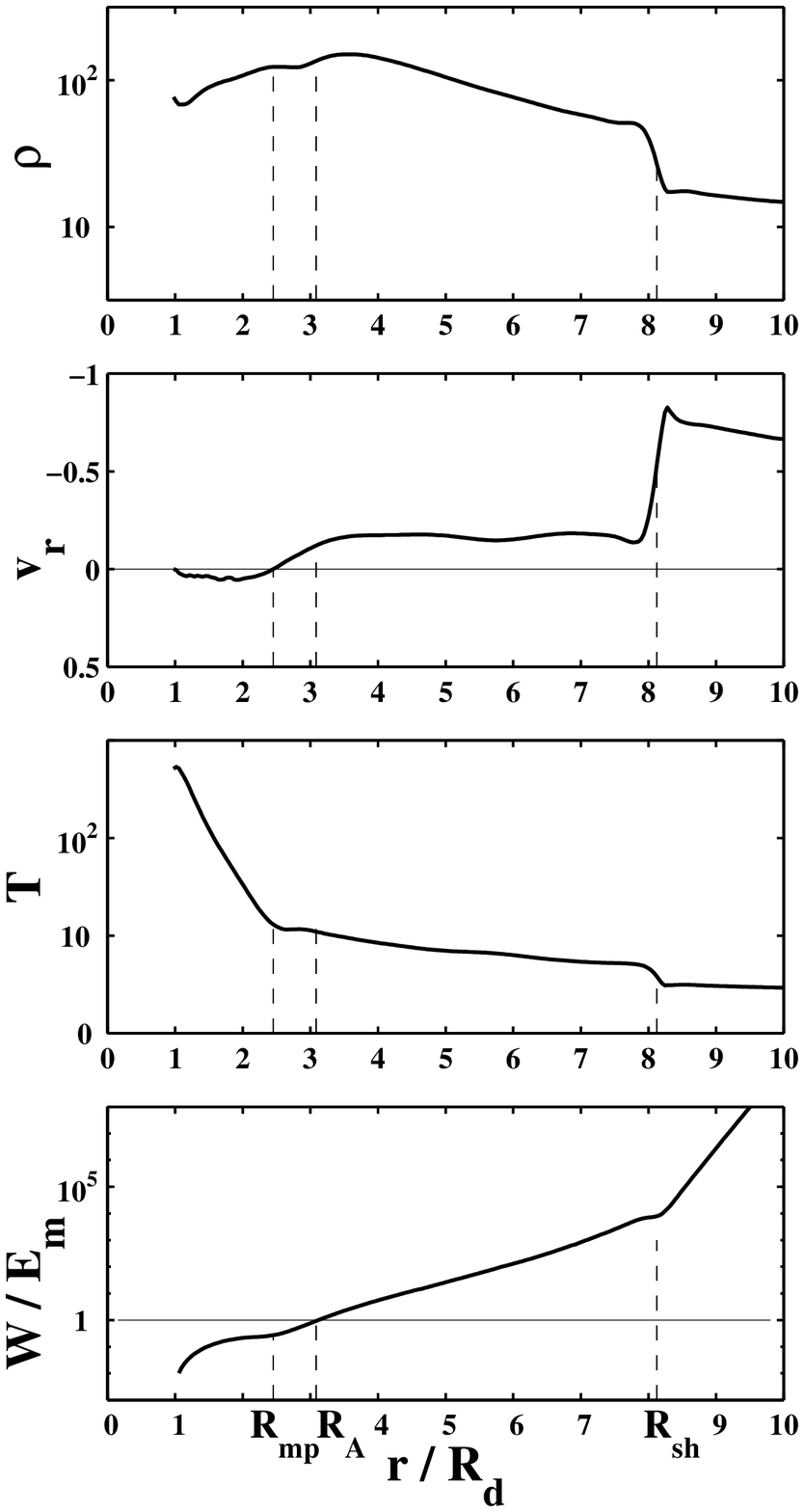}{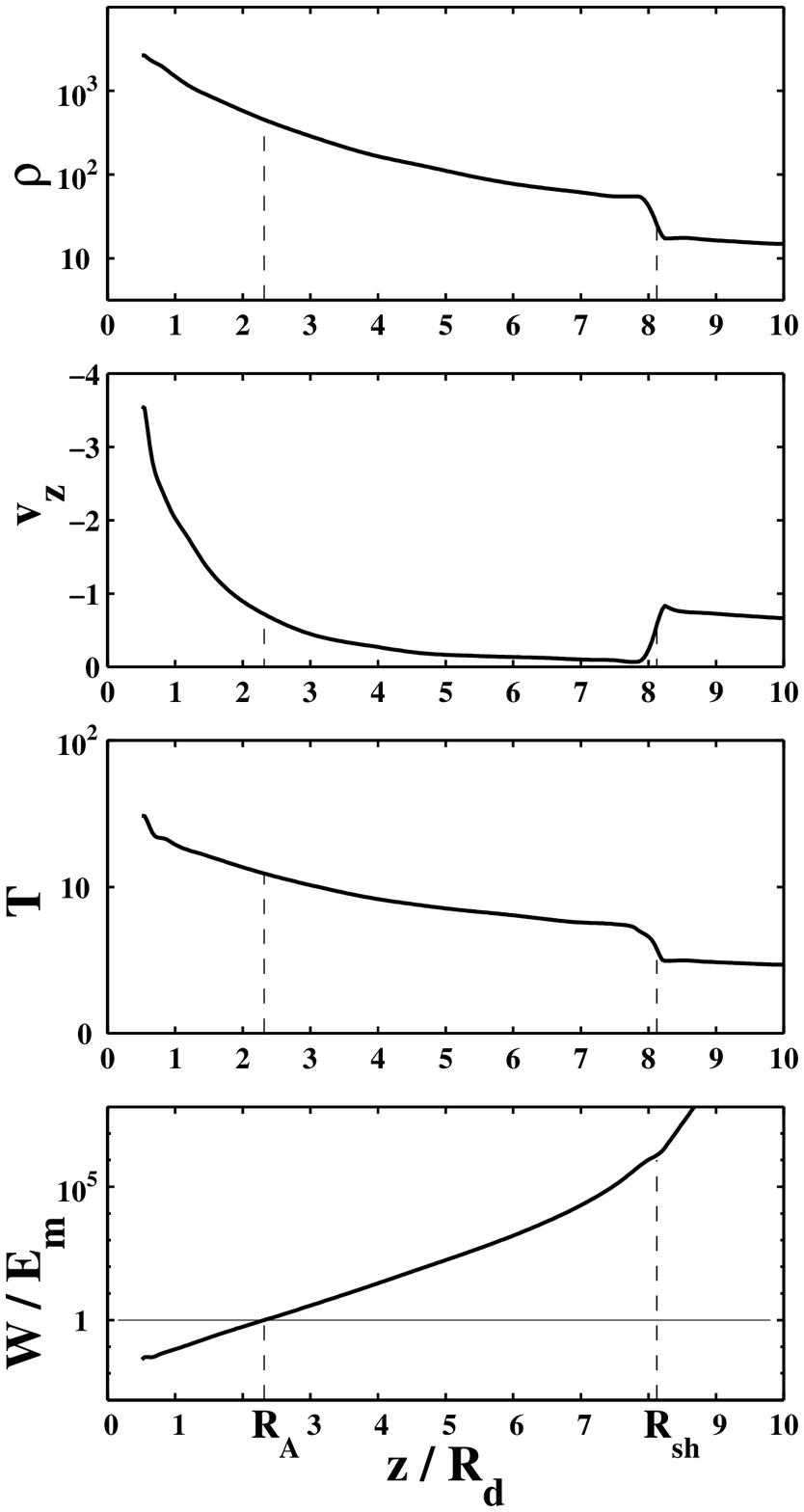}
\caption{
The figure shows the radial
variation of density $\rho$,
 radial velocity $v_r$, temperature $T$,
and ratio of total kinetic
energy-density of the matter
$W=\rho(\varepsilon + {\bf v}^2 /2)$ to
the magnetic energy--density $E_m={\bf H}^2/8 \pi$
 in the equatorial plane (left column) for
the acccretion flow presented at Figure~\ref{Figure 2}
$(t=3.5 t_{ff})$.
The positions of
the Alfv\'en radius $r=R_{A}$,
the magnetopause $r=R_{mp}$
 and the shock wave $r=R_{sh}$
are shown.
Right columns show the $z-$dependences
of the same
parameters along the $z-$axis.
The positions of the shock wave
$z=R_{sh}$ and Alfv\'en radius $z=R_A$ are shown.
}
\label{Figure 4}
\end{figure*}

\section{Results}

  This work investigates
spherical Bondi--type accretion of plasma
to a  star with
 an aligned dipole magnetic
field.
 The resistive MHD equations (5) -- (11)
were solved
using the method described in \S 2.5
and initial and boundary
conditions described in \S 2.2 and \S 2.3.

We performed simulations for
$12$ different  values of
$\beta\propto\dot M_B/\rvecmu^2$ (equation 26), and
magnetic diffusivities
$\tilde\eta_{m}$ (equation 30).
 The calculated flows in all runs are
similar to that shown in Figure 2.
   In \S~3.1, we describe
in detail the nature of the
flow for a representative run.
In \S~3.2 we analyze the stationarity of the
spherical accretion to a dipole.
  In \S~3.3, we show the dependence of
the flows on
 $\beta$ and $\tilde\eta_{m}$.
  In \S~3.4, sample runs of accretion to
a rotating star are presented.
  In
\S~3.5, we give a numerical application of
our results.

\begin{figure*}[t]
\epsscale{0.5}
\plotone{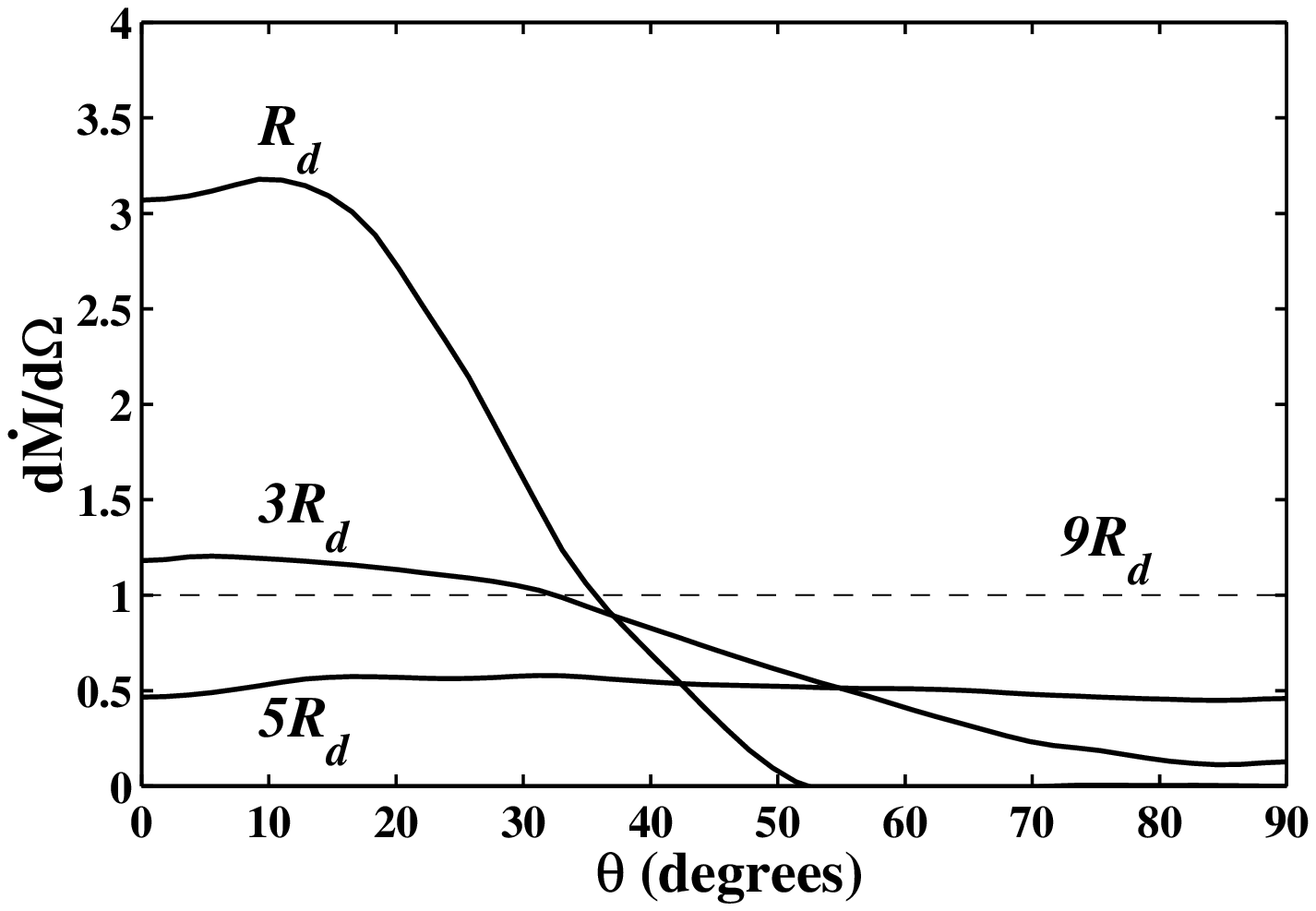}
\caption{
Differential mass
accretion rate per unit solid angle $d \dot{M}/d\Omega$
as a function of the
angle $\theta$ (with respect to
the $z-$axis)
for spheres with radii
$R_d$, $3R_d$, $5R_d$, and $9R_d$
for the flow presented in Figure 2.
  At large distances ($R=5R_d$) accretion
is almost spherically symmetric.
  Closer to
the dipole ($R=3R_d$),
 the flow becomes anisotropic.
 At very small distances
($R=R_d$) most of the matter
accretes to the poles
along a narrow cone of
half--angle $\theta_{1/2}
\approx 30^{\circ}$.
}
\label{Figure 5}
\vspace{1.5cm}
\epsscale{0.5}
\plotone{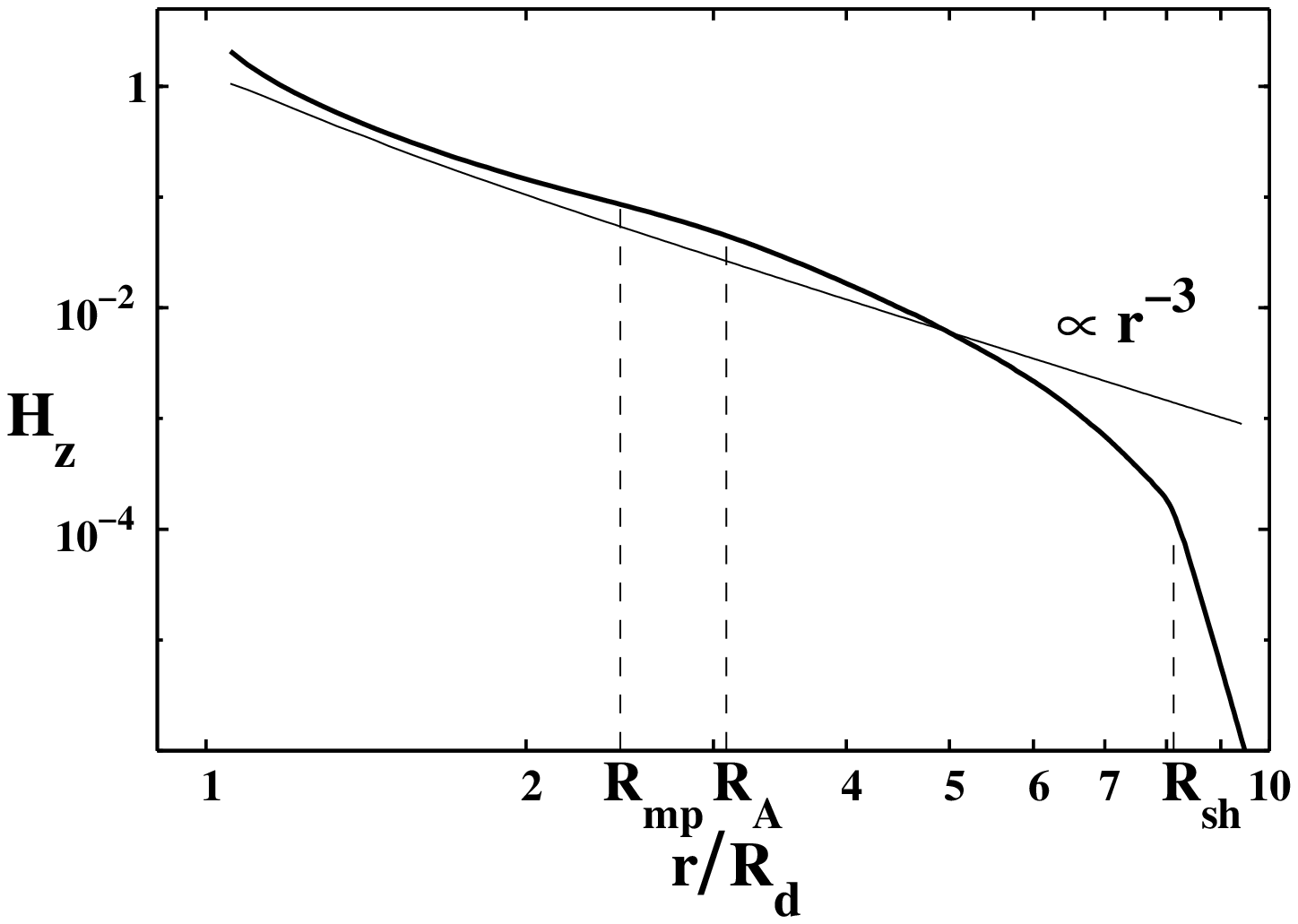}
\caption{
  Radial variation of
the vertical magnetic field $H_z$
in the equatorial plane.
The thin line shows
 the dependence of $H_z$ for
a vacuum dipole field.
The vertical dashed lines
indicate the positions of the shock wave
 $R_{sh}$, Alfv\'en radius
$R_A$, and magnetopause
radius $R_{mp}$.
}
\label{Figure 6}
\end{figure*}

\begin{figure*}[p]
\epsscale{0.475}
\plotone{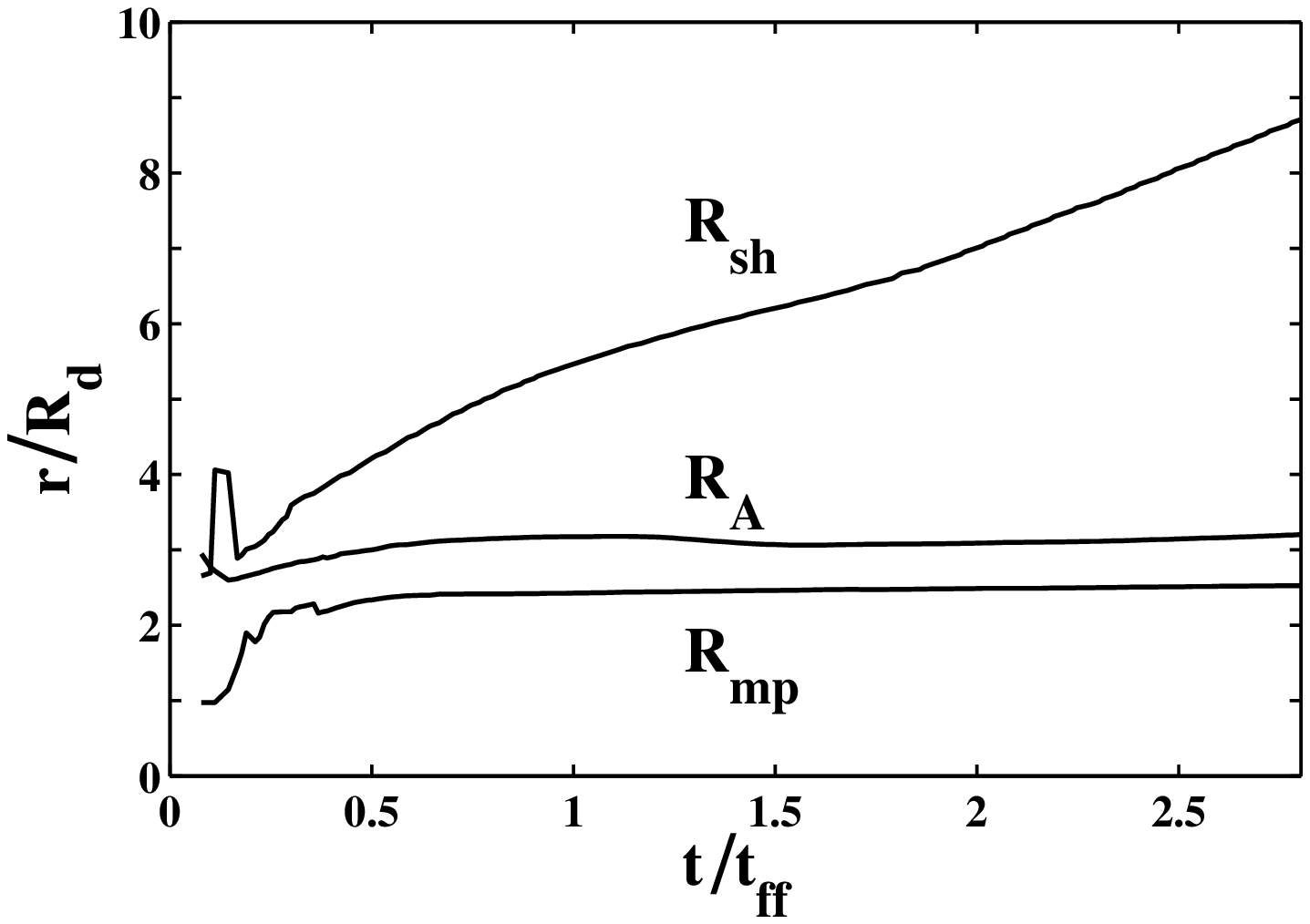}
\caption{
Temporal evolution of the shock wave
radius
$R_{sh}$,
Alfv\'en radius $R_A$, and
magnetopause radius
$R_{mp}$.
 Time is measured in units of
 $t_{ff}$ which is the
free--fall time from the distance $R_{max}$.
}
\label{Figure 7}
%
\vspace{1.cm}
\epsscale{0.5}
\plotone{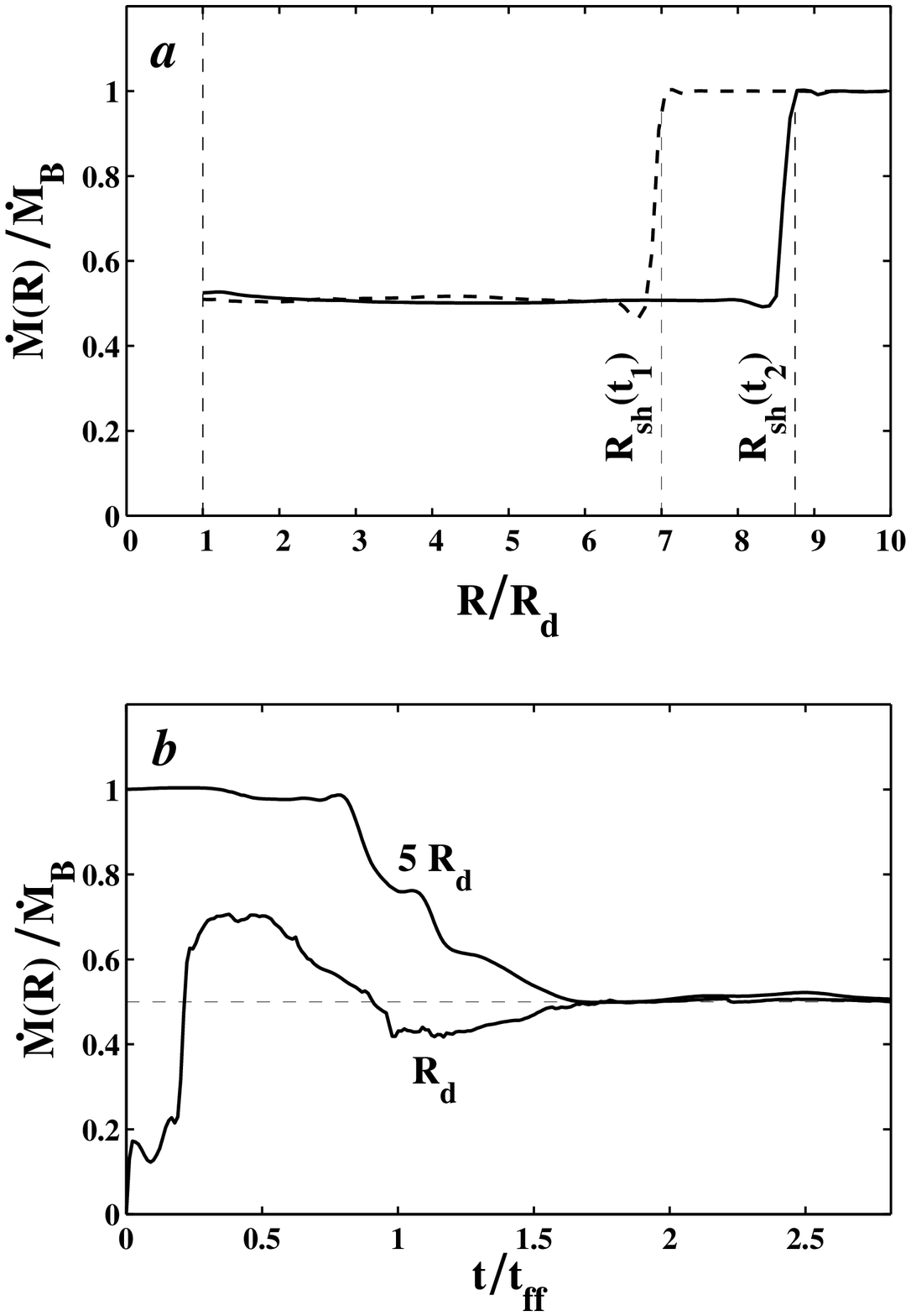}
\caption{
(a) Mass accretion rate $\dot{M}$ through
spheres of radii
 $R$ at $t_1=2 t_{ff}$
(dashed line) and
$t_2=2.8 t_{ff}$ (solid line).
  The shock wave expands from position
$R=R_{sh}(t_1)$
to position $R_{sh}(t_2)$.
Inside the shock the flow is subsonic
with accretion rate
$\dot{M}_{dip}\approx{}0.5 \dot{M}_B$.
(b)
Time evolution of mass--fluxes $\dot{M}(R)/\dot{M}_{B}$
throught spheres with radii
$R=R_d$ and $R=5R_d$.
}
\label{Figure 8}
\end{figure*}

\subsection{Illustrative Simulation Run}

Here we describe in detail a run with
$\beta=3.5 \times 10^{-7}$ and
 $\tilde\eta_{m}=10^{-5}$.
Simulations
were performed on a uniform grid with
$257{}\times{}257$ square cells.
  The size of the
computational region was
$10R_{d}\times{}10R_{d}$,
the accretor radius was
$0.5R_{d}$, where $R_d$ is the radius
of the current disk.
  We measure time
in units of the
free--fall time
from  $r=R_{max}$, $z=0$;  that is,
$t_{ff}=R_{max}^{3/2}/\sqrt{2GM}$.

  Spherical accretion to a
magnetic dipole is very different from that to
a non--magnetized star.
  Instead
of supersonic steady inflow, which is
observed in standard Bondi
 accretion,
a shock wave
 forms around the dipole.
  The
supersonic inflow outside
the shock becomes subsonic
inside it.
  In all cases we observe
that the shock wave
gradually expands outwards.
    Figure~\ref{Figure 2} shows
the main features of the flow
at  time $t\approx 2.5 t_{ff}$
when the shock has
moved to the distance
$R_{sh}=8.1 R_{d}$.
  We observe that for
$R > R_{sh}$ the flow is unperturbed
Bondi flow, whereas
inside the shock for $R < R_{sh}$
it is subsonic.
   Initially, the subsonic
 accretion to dipole is
spherically symmetric,
but closer to the dipole
it becomes
strongly anisotropic.
   Near the dipole matter moves along
the magnetic field lines and
accretes to the poles.
   Figure~\ref{Figure 3} shows the
inner subsonic region of the flow
in greater detail.
    The dashed line shows
the Alfv\'en surface,
which we determine as
the region where the
matter energy--density
$W=\rho(\varepsilon+{\bf v}^2\!/\!2)$
is equal to the magnetic
 energy--density
$E_m={\bf H}^2/(8\pi)$.
    The Alfv\'en
surface is ellipsoidal, with
radius
$R_A=3.1 R_d$ in the equatorial plane,
 and $R_A=2.3 R_d$
along the $z-$axis.
 Note, that the ``theoretically'' estimated
 Alfv\'en radius~(28)
is $R_A^{th}\approx{}2.6 R_d$.
A significant deviation from spherically
symmetric flow is observed
for $R\lesssim 2 R_A$, because
magnetic field starts to influence
the flow before it
reaches the Alfv\'en surface.
   Matter  in the equatorial plane
moves across the magnetic field lines,
decelerates and stops at a radius
$r=R_{mp}\approx 2.6 R_d$, which we term
the ``magnetopause radius.''
   There is a torus shaped
region -- a stagnation
 region -- which is avoided by accretting matter.
The flow velocities in this region are
negligible.
   The magnetopause region is located
 inside the Alfv\'en surface (see Figure~\ref{Figure 3}).
The accretion flow along the $z-$axis
is accelerated and becomes supersonic
at $z \approx 1.6 R_d$
(see inner dashed line at Figure~\ref{Figure 3}).

Figure~\ref{Figure 4} show the radial
and axial
variation of different parameters.
  One can see, that
the density $\rho$ is
larger in the subsonic
region
compared with the Bondi solution.
  The velocity
decreases by about a factor
of $4$ across the
shock wave, and the temperature
increases.
  The density in the magnetopause
region $R \leq R_{mp}$ is
lower than outside, while
the temperature
 is larger (see Figure~\ref{Figure 4}, left column).
  The bottom
panel of Figure~4 shows
the variation
 of ratio $W/E_m$.
  The radius where
$W/E_m=1$ is Alfv\'en radius.
 In the $z$-direction
(see Figure~\ref{Figure 4}, right column),
matter moves along the magnetic field lines
with the result that the
flow parameters change smoothly.

    Figure~\ref{Figure 5} shows the transition
from spherically
 symmetric
accretion outside the magnetosphere
to highly anisotropic accretion
along the polar columns within
the magnetosphere.
  The figure shows the matter flux
accreting through the
unit solid angle
$d\dot{M}
 /d\Omega$
at different
inclinations $\theta$ of this solid
angle relative to the $\pm z$ axis.
One can see that at
large distances $R=5 R_d$,
the accretion is almost
 spherically symmetric,
whereas at smaller
distances it becomes more
and more anisotropic.

Figure~\ref{Figure 6} shows that
the initial vacuum dipole
 magnetic field  (Figure~\ref{Figure 1}) is
strongly compressed
by the incoming Bondi flow.
  As a result of
interaction with Bondi flow,
the magnetic field has
a dipole dependence
only inside Alfv\'en radius,
$r \lesssim R_A$,
and it decreases faster
than the dipole
field for $R_A \leq r \leq R_{sh}$.
   This is a result
of an induced azimuthal shielding current
in the dipole's magnetosphere.
Thus distributed  current
has a sign opposite to that of the
 star's intrinsic azimuthal
current.
For $r \gtrsim R_{sh}$, the field
decreases dramatically with $r$
due to the shielding current.
 Thus, the magnetic field
at the outer boundary
of the simulation region is
negligible.

  Figure~\ref{Figure 7} shows that the
shock wave initially forms
close to the magnetosphere
at $R \sim R_A$ and
then gradually expands outward.
   However, the
Alfv\'en radius $R_A$ and magnetopause
radius $R_{mp}$ become steady
after $t \gtrsim t_{ff}/2$ and remain steady
thereafter.
   This means, that the
magnetosphere of
 the star reaches equilibrium with
the incoming matter
rapidly and this equilibrium
does not change
as a result of outward movement
of the shock wave.

\subsection{Stationarity of
 Accretion Flow to the Dipole}

   It is important to know whether
the calculated accretion flow
to the dipole is stationary or not.
We analyze this here.

\begin{figure*}[b]
\epsscale{0.5}
\plotone{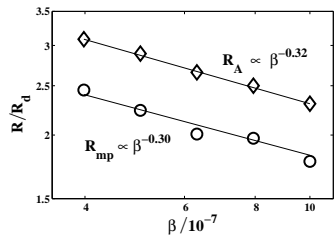}
\caption{
Dependence of the Alfv\'en radius $R_A$
and  magnetopause radius $R_{mp}$
on parameter $\beta \propto \dot{M}_B/\mu^2$
for magnetic diffusivity
 $\tilde{\eta}_m=10^{-5}$.
}
\label{Figure 9}
\end{figure*}

After the passage of the shock wave, the new
subsonic regime of accretion
forms around the dipole.
   The region of subsonic flow expands together
with the expanding shock wave.
   Here, we analyze
the stationarity of the flow in the
subsonic region.
  Figure~\ref{Figure 8}a shows the distribution
of matter fluxes through
spheres of different radii $\dot M (R)$
as a function of $R$
at $t = 2t_{ff}$ and
 $t = 2.8t_{ff}$.
One can see that
matter flux outside the shock wave
$R > R_{sh}$
is the Bondi rate $\dot M_B$, while inside
the shock wave it is significantly less.
   It is almost constant and equal
to the accretion rate to the dipole
$\dot M(R)\approx \dot M_{dip}\approx
0.5 \dot{M}_B$.

    Figure~\ref{Figure 8}b shows the matter fluxes through
fixed spheres located at radii $R=R_d$ and
$R=5 R_d$ as a function of time.
   The matter flux through the sphere of radius
$R=R_d$ corresponds to the matter flux
to the dipole.
    It decreases and goes to the
constant $\dot M=\dot M_{dip}$
 at $t>1.5 t_{ff}$.
   The matter flux
through the sphere $R=5 R_d$ is
initially the Bondi
rate, but after passage of the shock wave
it decreases  to $\dot M = \dot M_{dip}$
and does not change thereafter.
    Thus, Figures~\ref{Figure 8}a and \ref{Figure 8}b
demonstrate that the flow in the subsonic
region is stationary in both space and   time.
    The shock wave switches the
Bondi flow to a new flow with
stationary subsonic accretion
and smaller accretion rate.
   The local physical variables,
for example,  density $\rho$
 and velocity ${\bf v}$ are also
time independent.

 The  formation of an expanding shock wave
during accretion to a
 dipole results from
the fact that the gravitating center
with dipole  field  ``absorbs''
matter at a slower rate than the Bondi rate.
 The rate of accretion $\dot M_{dip}$
at given parameters of the Bondi flow is
determined by the physical parameters of
the dipole (see \S~3.3).

    An analogous situation was found in
investigations of hydrodynamical
accretion to a gravitating center.
   Kazhdan \& Lutskii (1977)
(see also Sakashita
1974; Sakashita \& Yokosawa 1974;
Kazhdan \& Murzina 1994)
investigated spherically
symmetric accretion flows
for conditions where
the matter flux through the
inner boundary (which is the
surface of the star) is less than
matter flux supplied at the outer boundary.
  They found a family of self--similar
solutions where the expanding
shock wave links the regions inside and outside
the shock wave.
   These regions
have different stationary matter fluxes
corresponding to matter fluxes at the
boundaries.
    Our simulations show similar behaviour.

    Here, we should point out that the shock wave
in our simulations is  a temporary phenomenon,
which  establishes a new regime of accretion
around the dipole.
    It appeared because
the external accretion rate is larger
than accretion rate which dipole can ``absorb.''
   A different situation was
considered by Ruffert (1994)
 who performed 3D hydrodynamical simulations of
Bondi accretion.
    He used  initial conditions where the
matter distribution had
constant density and zero velocity.
    In his simulations,
the initial matter flux is zero and
stationary
accretion was established
by  a rareaction wave
propagating outward from the
central  object.

    For fixed boundary conditions (supersonic
Bondi accretion  at the outer boundary)
the results of simulations
do not depend on initial conditions.
The dependence on boundary conditions will
be investigated separately.

    It is of interest to know
how far  outward the the shock wave will
propagate.
  Note that
the Bondi flow is supersonic out
to some distance and is subsonic at larger distances.
We expect that after reaching the subsonic area,
 the shock wave will ``dissolve'' and the flow will be
purely subsonic.
    Thus, the shock wave may
be only a temporary phenomenon
which results from the initial conditions
of our simulations.
  From the other side, if the flow is supersonic up
to very large distances, then the shock wave
expansion  may be stopped by the physical
structure of the astrophysical system.
   For example, in the wind-fed pulsars,
the shock movement would stop at
a radius of the order
of the binary separation.

\begin{figure*}[t]
\epsscale{0.5}
\plottwo{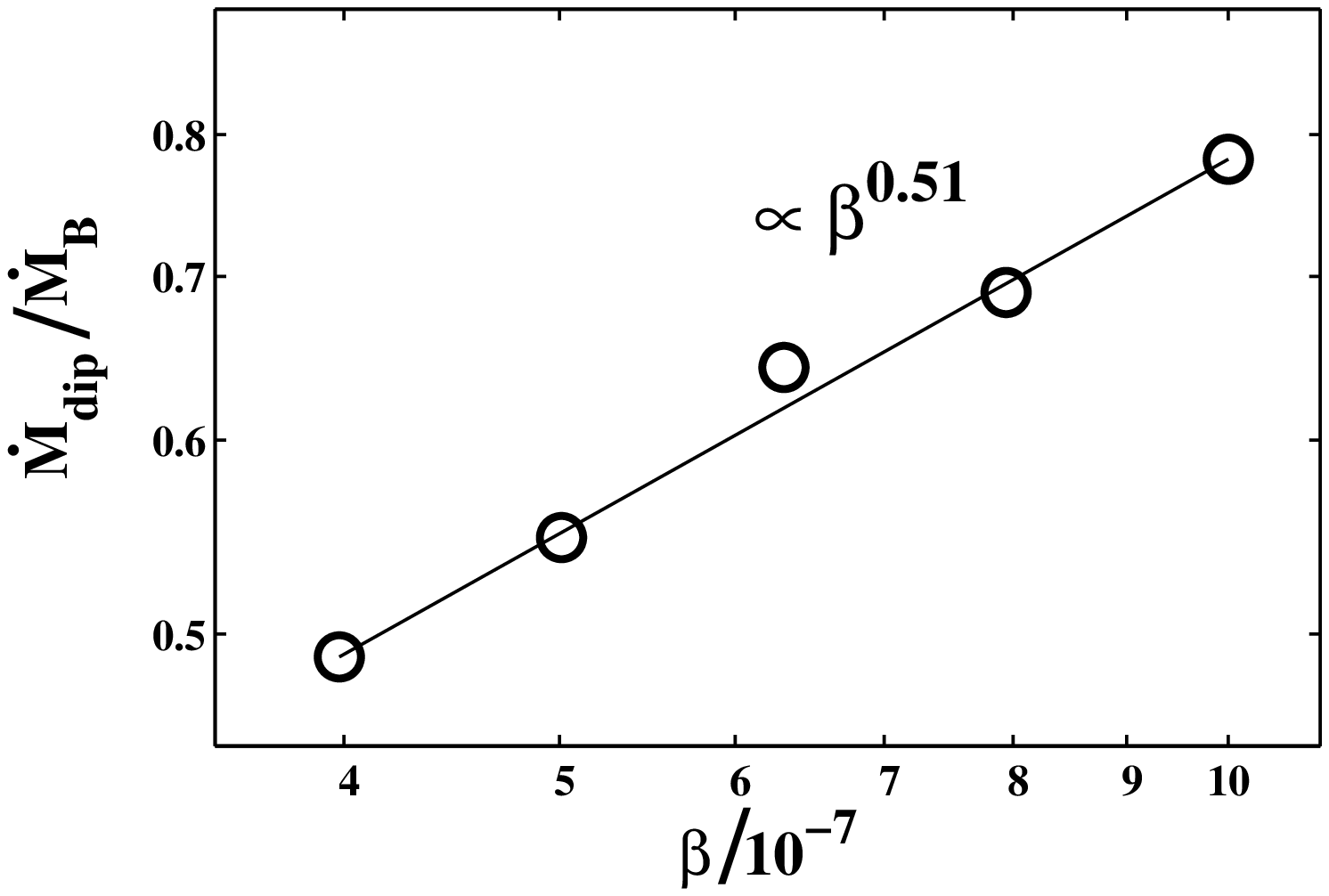}{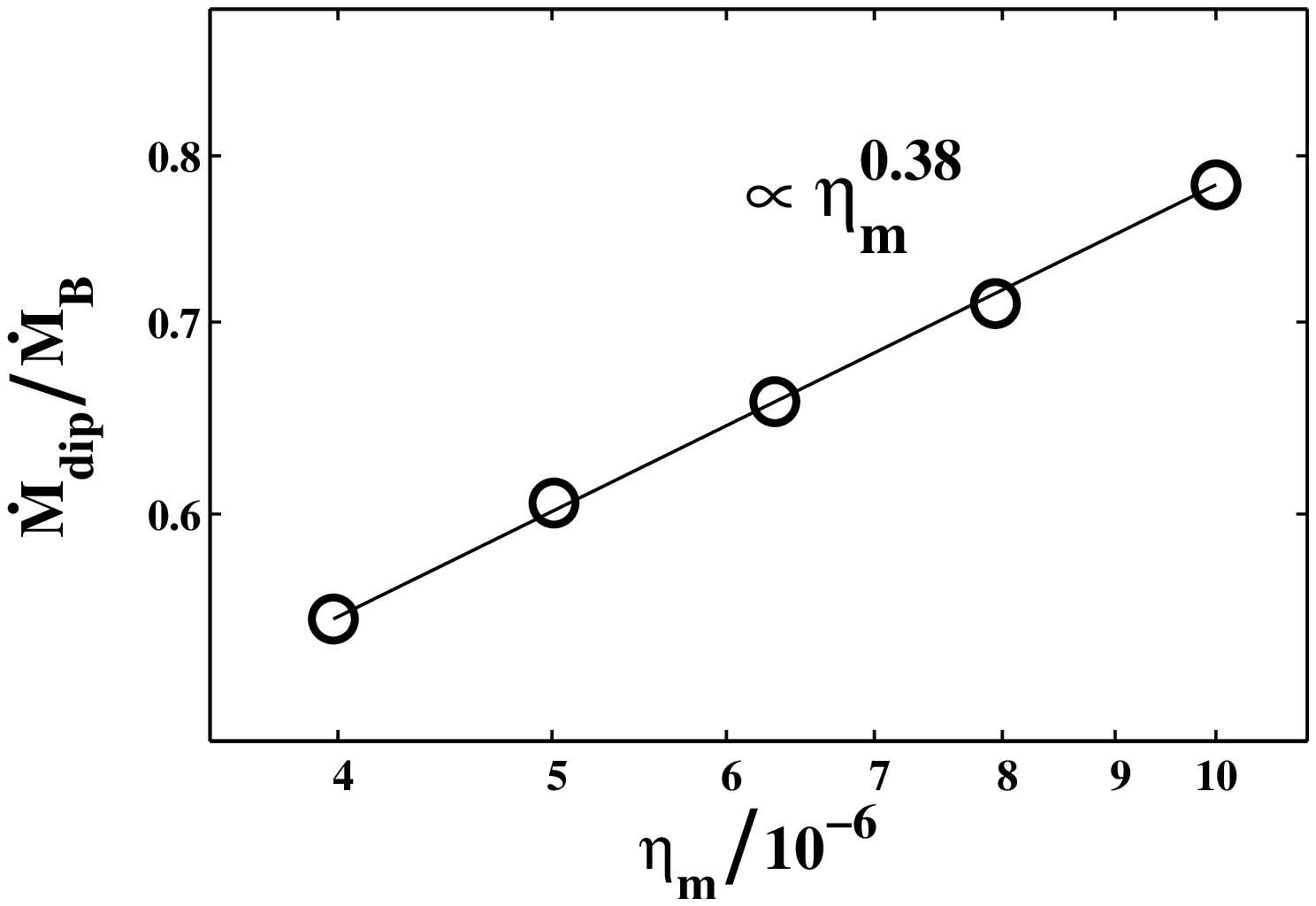}
\caption{
Dependence of accretion
rate to the dipole
$\dot M_{dip}$, measured in Bondi
accretion rate $\dot M_B$,
on $\beta$ (at
 $\tilde{\eta}_m=10^{-5}$, left frame)
and on  $\tilde{\eta}_m$
(at $\beta=10^{-6}$, right frame).
}
\label{Figure 10}
\end{figure*}

\subsection{Dependence of
 the Accretion Flow
on $\beta \propto
{\dot M_B/\mu^2}$ and $\tilde\eta_m$}

 We first analyze the dependence of the
flow on the external
accretion
rate $\dot M_B$ and
 the star's magnetic moment $\rvecmu$.
  As discussed in \S~2.4,
these quantities are coupled so that
the investigated physical model
depends only on the ratio
$\beta{}\propto{}{\dot M_B}/{\rvecmu^2}$.
  Each simulation run takes considerable
time and for this reason
we adopted the following  procedure
for deriving the
dependence on $\beta$.
  We start from the conditions
of the simulation run
of with $\beta=10^{-6}$ at time
$t=2t_{ff}$,
and then change $\beta$ by a factor
of $10^{n/10}$ in a sequence of $5$
independent simulations ($n=1\ldots 5$).
  These simulations were performed
up to $t=5t_{ff}$.
   Fluctuations
connected with
the readjustment of the flow
are damped by this time.
   We then measured the
radius of the magnetopause
$R_{mp}$ and the Alfven radius
$R_A$ for new values
of $\beta$.
  The  Alfv\'en radius in the
equatorial plane is found to
have a power law dependence,
\begin{equation}
\label{beta-alfven}
   R_{A}(\beta)\approx
   2.3{}R_{d}
   \left(\frac{\beta}
   {10^{-6}}\right)^{-k_\beta}{~,}
\end{equation}
where  $k_\beta\approx
0.32\pm 0.03$ $(\sim 2/7\approx 0.286)$.
The equatorial magnetopause
 radius is found to be proportional
to the Alfv\'en radius,
$ R_{mp}(\beta)\approx 0.026 R_{d}
+ 0.756 R_{A}(\beta)$.
Figure~\ref{Figure 9} shows the observed dependences.
   The Alfv\'en radius is found
to have a weak dependence on
magnetic diffusivity $\eta_m$,
$R_A\sim \tilde\eta_m ^{-k_\eta}$,
where $k_\eta\approx 0.075$.

    Figure~\ref{Figure 10} shows that the stationary accretion
rate to the dipole
${\dot M}_{dip}$
also depends on $\beta$.
  We find that $\dot{M}_{dip}$ is
 always smaller than the
Bondi accretion rate $\dot{M}_{B}$.
  The dependence  found is
\begin{equation}
\label{dotM_bet}
{\dot M}_{dip} \approx
0.78{\dot M}_{B}
\left(\frac{\beta}{10^{-6}}
\right)^{m_{\beta}}{~,}
\end{equation}
for $\beta \leq 10^{-6}$,
where $m_{\beta}\approx 0.51$.

\begin{figure*}[p]
\epsscale{0.6}
\plotone{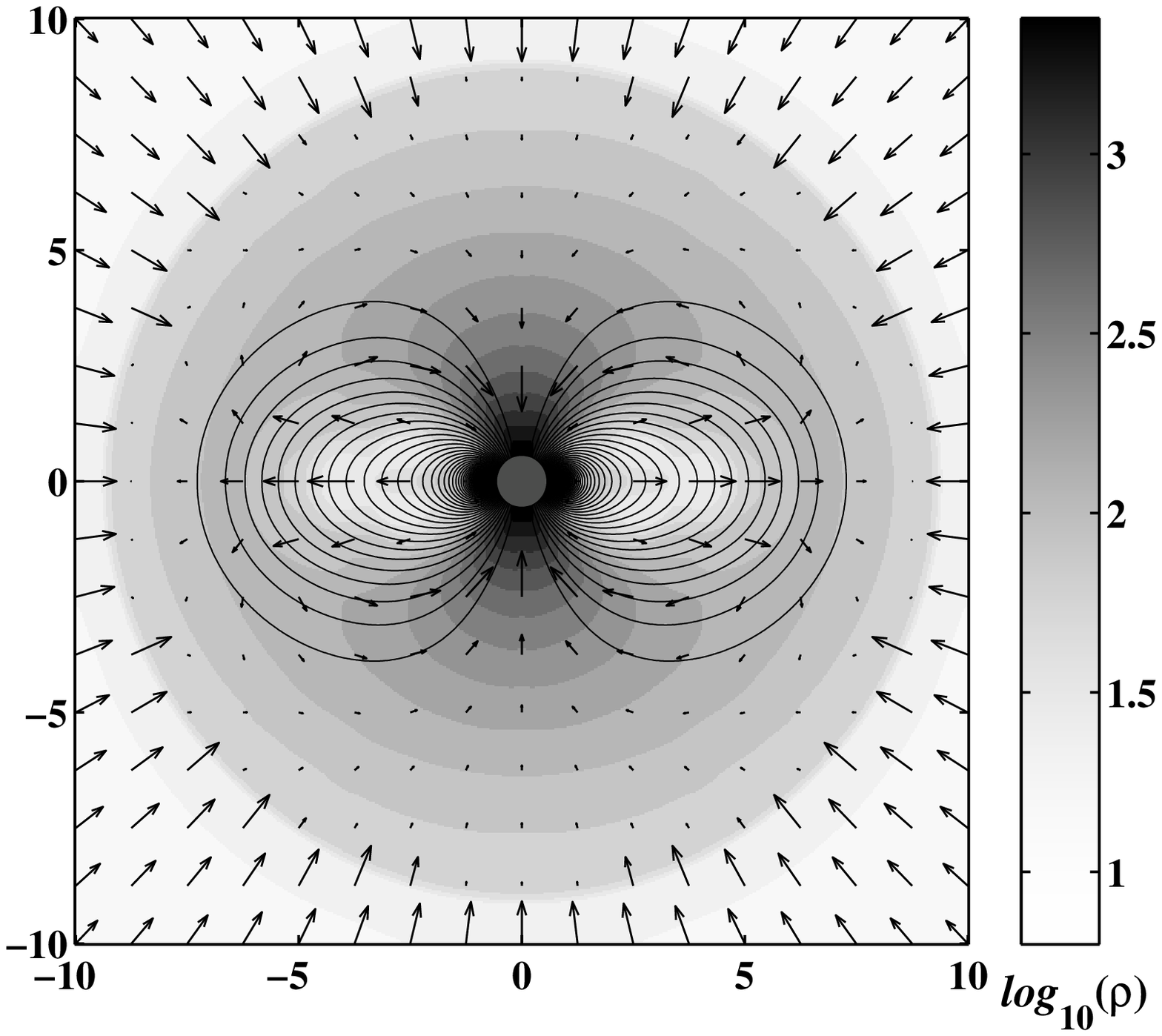}
\caption{
The figure shows the
accretion flow
to a rapidly rotating star
(see \S 3.3) at  time
$t=2.5t_{ff}$ .
  The gray--scale background
represents the density
and the solid lines the
poloidal magnetic field.
   The length of arrows is
proportional to the flow speed.
}
\label{Figure 11}
\vspace{0.5cm}
\epsscale{0.6}
\plotone{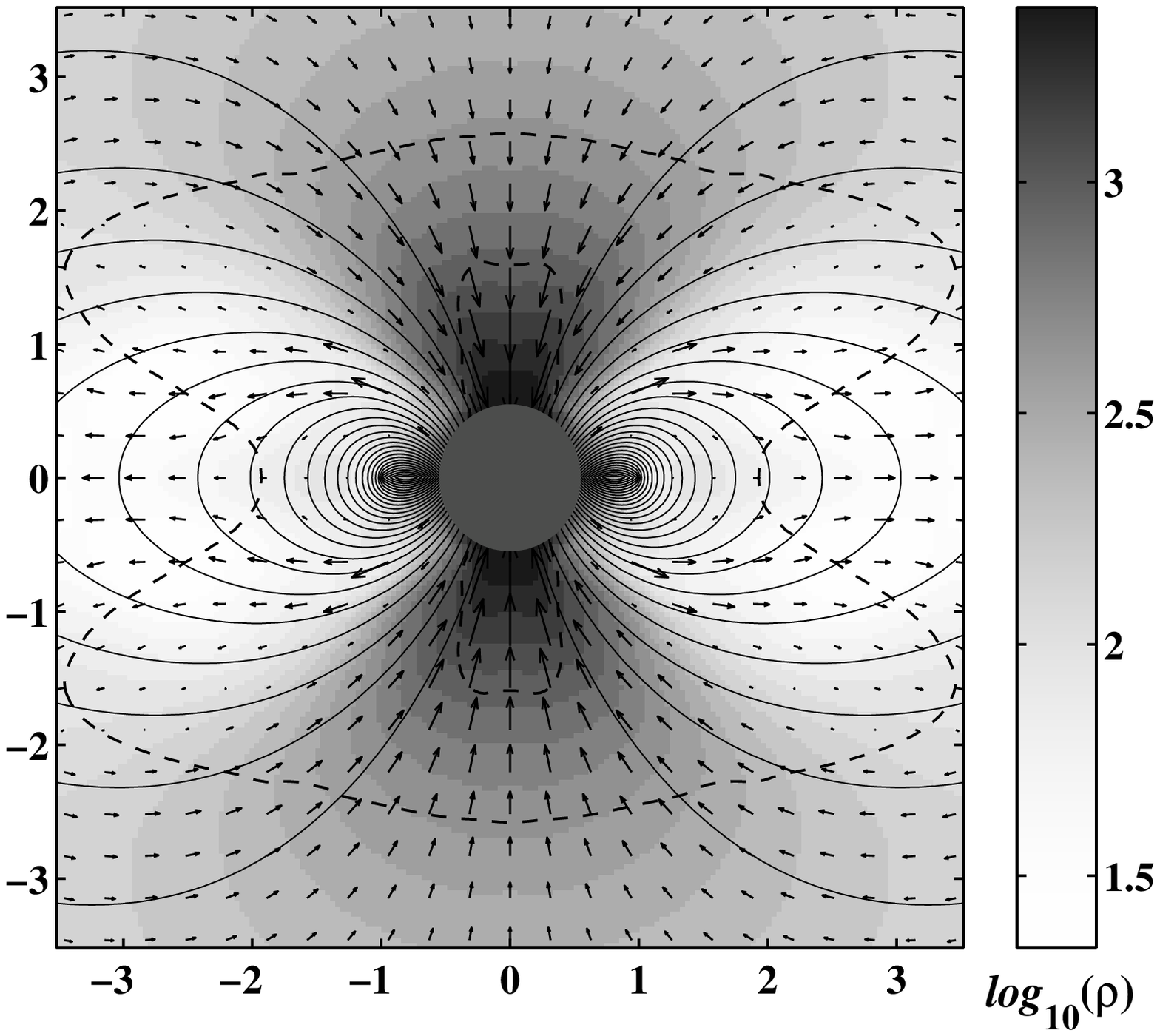}
\caption{
Enlarged view of the
accretion flow to a
rapidly rotating star with dipole
field (Figure~\ref{Figure 11}).
   The outer
dashed line represents
the Alfv\'en surface.
A inner sonic surface is
indicated by the
dashed
line in the region of polar columns.
    The outflow in the equatorial plane starts
from the region of the corotation radius
$R_{cr} = 2{}R_{d}$ inside
magnetosphere.
}
\label{Figure 12}
\end{figure*}

\begin{figure*}[t]
\epsscale{0.5}
\plotone{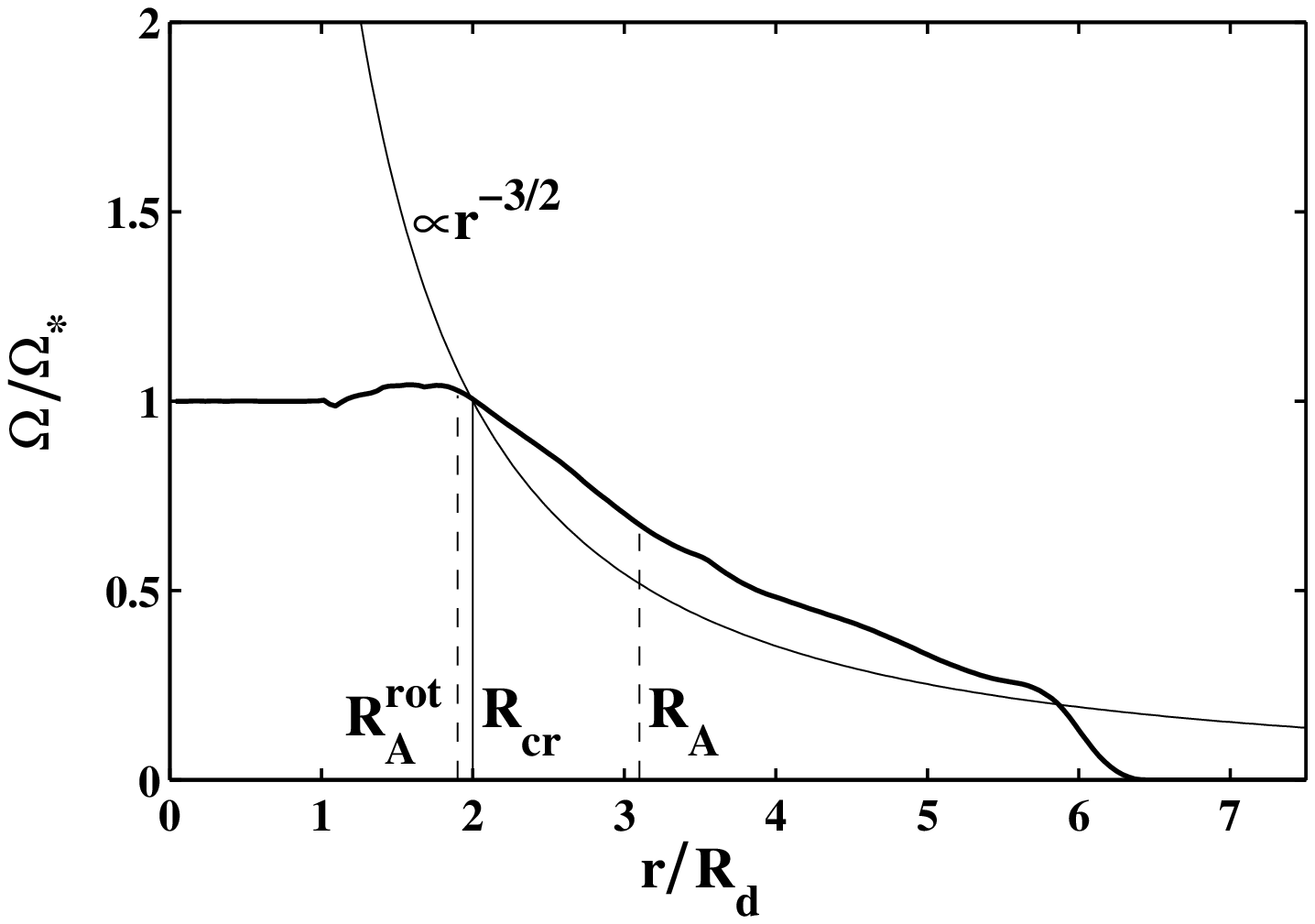}
\caption{
The radial variation of the angular
velocity of the
 matter $\Omega(r) = v_\phi(r)/r$
in the equatorial plane measured in
units of angular velocity of the
disk $\Omega_\star$ (solid line).
   The thin solid line shows the
radial dependence of the
Keplerian
angular velocity.
The positions of the corotation radius
$R_{cr}$ and $R_A^{rot}$  are shown.
Also, the position of the Alfv\'en radius for
the corresponding non--rotating system
$R_{A}$ is marked.
}
\label{Figure 13}
\end{figure*}

Here, we recall that
$\beta\propto{}\dot{M}_{B}/\rvecmu^2$
and in equation (23) for $\dot{M}_{B}$
all quantities except $\rho_\infty$
are fixed in our simulations.
   Thus, the accretion rate to the dipole
depends on the density of
surrounding matter as
$\dot{M}_{dip}\propto\rho_\infty^{3/2}$
and on
the star's magnetic moment as
$\dot{M}_{dip}\propto{}1/{|\rvecmu|}$.
The last dependence can be explained by
the fact that at larger
 $\beta$,
the Alfv\'en radius is smaller
so that the opening
angle of the accretion
columns is larger.
As a result, matter can
more readily
flow  into the gravitating  center.
At  $\beta > 3\cdot{}10^{-6}$, the
dependence is different.
${\dot M}_{dip}$ increases
more gradually and approaches the
critical Bondi accretion
rate $\dot M_B$. Here,
 we observe
stationary Bondi flow similar to
 that observed in simulations
to non--magnetized gravitating
object (Ruffert~1994).

  Figure~\ref{Figure 10} shows the dependence of
the accretion rate on the
magnetic diffusivity
\begin{equation}
{\dot M}_{dip}\approx
0.78{\dot M}_{B}
\left(\frac{\tilde\eta_{m}}{10^{-5}}
\right)^{m_{\eta}}{~,}
\end{equation}
for $\tilde\eta_m \leq 10^{-5}$,
$m_{\eta} \approx 0.38$.
This dependence means that matter
accretes more readily
at larger diffusivity as
expected.

The half--opening angle $\theta_{1/2}$
of the accretion funnel at $r=R_d$ (see Figure~\ref{Figure 5})
decreases as $\tilde\eta_{m}$ decreases. We find
$\theta_{1/2}\propto{}(\tilde\eta_{m})^{0.26}$.

At  higher difusivities
$\tilde \eta_{m} > 10^{-5}$, the dependence (33)
becomes smoother, $\dot M_{dip}\rightarrow \dot M_B$
and we observe steady accretion at the Bondi rate.

\subsection {Accretion to a
 Rotating Dipole}

    We also investigated
cases of accretion to
a {\it rotating} star with an aligned
dipole magnetic field.
   To create the rotating dipole, we
rigidly rotated the current
disk $0\le r \le R_d$ which is a
part of  boundary condition
at $z=0$ as discussed in \S 2.3.
   The disk radius $R_d$ is
in effect the radius of the star.
   We discuss two cases, one
with slow rotation,
$\Omega_\star=0.1 \Omega_{K}(R_d)$, and
the other
with fast rotation,
$\Omega_\star=0.35 \Omega_{K}(R_d)$,
where $\Omega_\star$ is the star's
angular rotation rate and
$\Omega_{K}(R_d)=
\sqrt{GM/R_d^3}$ is the Keplerian
angular velocity at
the edge of the disk $r=R_d$.

    We observed that
in the case of slow rotation
the general behavior of  accretion
flow is similar to that
for the non--rotating case.
The corotation radius $R_{cr}$, where $\Omega_K(R_{cr})
= \Omega_\star$ or $R_{cr} =
(GM/\Omega_\star^2)^{1/3}$, is significantly
larger than the Alfv\'en radius $R_A$
for a non-rotating dipole.

    As in the non-rotating case, the shock wave
forms and propagates outwards,
while accretion to the
 dipole is subsonic and steady.
However, a new feature
appears:  The stagnation region
mentioned earlier
rotates rigidly with the angular velocity
of the star.
   We find that the
limit of slow rotation is valid
for $R_{cr}> R_A$.
In this case the linear
velocity of rotation at the outer
edge of magnetosphere
($R=R_A$) is smaller than
the Keplerian velocity.
    In cases of slow rotation,
accretion to the dipole is
steady but
with an
 accretion rate $\dot M_{dip}^{rot}$
which is smaller
 than the corresponding value
for a non-rotating star, $\dot M_{dip}$.

    The second case we discuss is
a rapidly rotating dipole,
where the corotation radius
is smaller than the
Alfv\'en radius
for the corresponding system with a
non--rotating star,
$R_{cr}< R_A$.
   In this case the outer equatorial region
 of the rotating magnetosphere,
 $R_{cr}< r < R_A$,
 has azimuthal velocities in excess
 of Keplerian velocity.
Matter moves outwards
in a wide ``belt''-- like region
 around the equatorial plane.
     Figures~\ref{Figure 11} and \ref{Figure 12} show
the simulation results
for
$\Omega_\star=0.35 \Omega_K$ when the
corotation radius is
$R_{cr}=2.0R_d$, and
the  Alfv\'en radius in the equatorial plane is
 $R_A^{rot}=1.9R_d$.
     One can see, that magnetic field
lines are elongated in
equatorial direction by outflowing
matter.
The strongest outward
 acceleration is in the equatorial plane
where the centrifugal force is largest.
However, in the region
 of Alfv\'en surface (see Figure~\ref{Figure 12}),
an essential  acceleration is observed
along the magnetic field lines.
 This acceleration
appears to determine
the unusual shape of the
Alfv\'en surface (see Figure~\ref{Figure 12}).
Also, an essential outflow may
occur above and below the equator.

   At larger radial distances,
the outflowing matter encounters
the incoming Bondi flow
and turns to the direction
of poles.
    The magnetic field lines
are elongated in the $r$-direction.
   The Alfv\'en radius in the direction
of the poles has a value similar
to that in the non-rotating case.
    However, in the equatorial
plane the Alfv\'en radius
is significantly smaller ($R_A^{rot} \approx 1.9R_d$)
than its non-rotating value
($R_A \approx 3.1R_d$).
  Figure~\ref{Figure 13} shows that the magnetosphere inside
$r< R_A^{rot}$ rotates
rigidly.
    The angular velocity decreases gradually for
$r> R_A^{rot}$.
    The Figure
also shows that the
angular velocity of matter is larger than
Keplerian  in the outer parts of the
magnetosphere beyond the Alfv\'en
radius.
In this region, the centrifugal force
expells matter outward forming
``propeller''~-- like outflows.
   The ``propeller'' outflows
are predicted to occur in
the magnetosphere of
a rapidly rotating magnetized
star as first discussed by
Illarionov and Sunyaev (1975).
   The theory of such outflows
has received renewed interest
for the case of disk accretion
to rotating magnetized stars
 (Li \& Wickramasinghe 1997;
Lovelace, Romanova,
and Bisnovatyi-Kogan 1998).
    A systematic study
of spherical accretion to a
rotating star with an
 aligned dipole magnetic field
is in preparation
 (Toropin et al.~1998).

\subsection{Astrophysical Example}

Here, we present an application of
our simulation results in terms of
the physical quantities.
    We consider Bondi
accretion to a non-rotating magnetized
protostar
with  mass
$M=1~M_{\sun}=2 \times 10^{33}$~g
and  radius
$R=10^{11}$~cm.
    We use our
simulation run with $\beta=10^{-6}$
and $\tilde{\eta}_{m}=10^{-5}$,
which is close to the case
discussed in \S 3.1.
 We take the radius of the current disk
to be equal to the radius of the star,
$R_d=R$.
  From equation (24), the Bondi radius
is $R_B=50\sqrt{2}R
\approx~7.1 \times 10^{12}$~cm.
  The sound speed of the matter at
infinity from equation (26) is
$c_{\infty}=\sqrt{GM/R_B}
\approx{4.3}\times{10^6}$~cm/s.
This corresponds to a temperature
$T_{\infty} \approx 8 \times 10^4$ K for
a hydrogen plasma.

We assume that the magnetic field at
the star is $H_0=100$ G.
Then, according to equation (16),
magnetic moment of the star is
$\mu=1.4 \times 10^{34}$ ${\rm G}~{\rm cm}^3$.
  From equations (26),
the matter density
at infinity is
\begin{equation}
   \rho_{\infty}=
{\gamma\over{c_{\infty}}^2}~
    {H_0^2\over{8 \pi}}~ \beta~.
\end{equation}
For these parameters,  the
density at infinity is
$\rho_{\infty}
\approx 3 \times 10^{-17}~$g$/$cm$^3$
or a particle number density
$n_{\infty}=1.8 \times 10^7$1/cm$^{3}$.
The Bondi accretion rate from
equation (23)
is
$\dot{M}_B=8.3 \times 10^{-10}~
{\rm M}_{\sun}/{\rm yr}$.

Our code has finite magnetic
diffusivity $\eta_m$.
Here, we estimate the magnetic
 Reynolds
number $Re_m$.
At a distance $R$ from the origin
we have
\begin{equation}
   Re_m={{R ~v_{ff}}\over \eta_m},
\end{equation}
where $v_{ff}={\sqrt{2GM/R}}$ is
the free-fall speed.
  Using the definition of
dimensionless magnetic diffusivity, we get
$\eta_m=\tilde{\eta}_m R_B V_{A0}=
\sqrt{2GMR_B/{\gamma\beta}}$.
   Finally, we obtain
\begin{eqnarray}
\nonumber Re_m={{(\gamma\beta)^{1\over 2}}
\over{\tilde{\eta}_m}}
{\left({R\over R_B}\right)^{1\over 2}} =\\
={1.67~\left({R\over R_d}\right)^{1\over 2}}
 {\left({\beta\over
10^{-6}}\right)^{1\over 2}}
\left({10^{-5}\over \tilde{\eta}_m}\right)~.
\end{eqnarray}
At the distance of magnetopause,
 $R = R_{mp}\sim 3 R_d$,
we get $Re_m \sim 2.9$.

\section{Conclusions}

    We have developed a method
for MHD simulation of
spherical Bondi-type
 accretion flow to a rotating star
with an aligned dipole magnetic
field.
    Using this method
we have made a detailed study of the accretion
to a non--rotating star for
different accretion rates,
stellar magnetic moments, and
magnetic diffusivities.
  We also
include an illustrative case of accretion
to a rapidly rotating star.
  The simulation study confirms
some of the predictions of the analytical
models
(Davidson \& Ostriker 1973; Lamb et al. 1973;
Arons \& Lea 1976).
However, the simulated flows
show a different behavior from the models in
important respect summarized below.

  Our results for accretion to
a non--rotating star agree
qualitatively with some of the early
theoretical predicutions.   In particular,
(1) A shock wave forms around the dipole
   which acts as an obstacle
 for the accreting matter;
(2) A closed inner magnetosphere forms
where the magnetic energy-density
is larger than the matter energy density;
(3) The outer dipole magnetic field is
strongly compressed
 by the incoming matter.
(4) The flow is spherically symmetric at
large distances, but becomes  anisotropic
near and within the Alfv\'en surface.
Closer to the star the accretion flow
becomes highly anisotropic.
Matter moves along the polar magnetic field lines
forming funnel flows (Davidson \& Ostriker 1973).
(5) The Alfv\'en radius varies
with $\beta \propto \dot{M}_B/\mu^2$
as $R_A \sim \beta^{-0.3}$, which is
close to theoretical prediction
$R_A \propto \beta^{-2/7}$ (Davidson \& Ostriker 1973).

The new features observed in our
simulations of accretion
to a non-rotating star include the following:

(1) We observe
that the shock wave
which initially forms
around the magnetosphere
is {\it not} stationary
but rather expands outward
in all of our simulation runs.
This is different from  the theoretical
models which {\it assume} a stationary
or standing shock wave (Arons \& Lea 1976);
(2) A
{\it new stationary regime of subsonic
accretion}
forms around the star with
dipole magnetic field;
(3) A star
accretes matter only at specific $\dot{M}_{dip}$
rate which is less than
the Bondi rate $\dot{M}_B$.
That is,   $\dot M_{dip}=k~\dot M_{B}$
with $k< 1$;
(4) This accretion rate $\dot M_{dip}$
is smaller when
 $\beta\propto \dot M_B/\mu^2$ is smaller,
that is, when the star's magnetic
field  is larger.
Also, $\dot{M}_{dip}$ increases as
the magnetic diffusivity $\eta_m$ increases.

    We are presently making a systematic
study of accretion to a rotating star with
dipole field.  In this work we give only
sample results which illustrate the new
behavior resulting from the star's rotation.
 Accretion to a slowly rotating star,
where the corotation
radius $R_{cr} \equiv (Gm/\Omega_*^2)^{1/3}$
 is significantly larger than the Alfv\'en radius $R_A$
is similar to accretion to a
non-rotating star.
 However, the rate of accretion
$\dot M_{dip}$ is
smaller than in the corresponding
non-rotating case.
    For a rapidly rotating star, where
$R_{cr}< R_A$, ``propeller''
outflows form in the outer
 parts of magnetosphere
 and outside magnetosphere as proposed by
Illarionov and Sunyaev (1975).
These outflows result in a major change
in  accretion flow and field configuration.

\acknowledgments
{
The authors
 thank Prof. Bisnovatyi-Kogan
for many valuable discussions.
   This work
was made possible in part by
Grant No. RP1-173 of the U.S.
Civilian R\&D Foundation for the
Independent States
of the Former Soviet Union.
   Also, this work was supported in part by
NSF grant AST-9320068.
YT and VMC were supported in part
by the Russian Federal
Program ``Astronomy''
(subdivision ``Numerical
 Astrophisics'') and by INTAS
grant 93-93-EXT.
The work of
RVEL was also supported in part by NASA
grant NAG5 6311.}

\end{document}